\documentclass[aps,onecolumn]{revtex4}
\usepackage{graphicx}
\usepackage{amsmath}

\begin{document}

\title{Cavity approach to sphere packing in Hamming space}

\author{A.~Ramezanpour}
\affiliation{Physics Department and Center for Computational Sciences, Politecnico di Torino, Corso Duca degli Abruzzi 24, 10129 Torino, Italy}
\author{R.~Zecchina}
\affiliation{Physics Department and Center for Computational Sciences, Politecnico di Torino, Corso Duca degli Abruzzi 24, 10129 Torino, Italy}
\affiliation{Human Genetics Foundation, Torino, via Nizza 52, 10126 Torino, Italy}
\affiliation{Collegio Carlo Alberto, Via Real Collegio 30, 10024 Moncalieri, Italy}

\date{\today}

\begin{abstract}
In this paper we study the  hard sphere packing problem in the Hamming space by the cavity method.  We show that both the replica symmetric and the replica symmetry breaking approximations give maximum rates of packing that are asymptotically the same as the lower bound of Gilbert and Varshamov. Consistently with known numerical results, the  replica symmetric equations also suggest a crystalline solution, where for even diameters the spheres are more likely to be found in one of the subspaces (even or odd) of the Hamming space.  These crystalline packings can be generated by a recursive algorithm which finds maximum packings in an ultra-metric space. Finally, we design a message passing algorithm based on the cavity equations to find dense packings of hard spheres.  Known maximum packings  are reproduced efficiently in non trivial ranges of dimensions and number of spheres.
\end{abstract}


\maketitle

\section{Introduction}\label{0}
The problem of packing rigid objects, and spheres in particular,  is a fundamental problem which appears across  disciplines \cite{TS-rmp-2010,PZ-rmp-2010}. 
In general, the objects could have arbitrary shapes and the ambient space can be an abstract space  $\Lambda$.  
Given the space and the objects, the main question is that of finding the densest packings.

In coding theory one is interested in finding an optimal representation of $N$ symbols in binary strings of length $n$, 
that is $\Lambda=\{0,1\}^n$ is the Hamming space of dimension $n$. 
This optimal coding contains as many as possible symbols and ensures that after transmitting 
through a noisy channel, which at most flips $\frac{d-1}{2}$ variables, one can recover the original messages.
The ratio between the characteristic length of the symbols $l_c\equiv \log_2 N$ and length of the transmitted strings
$n$ defines the rate of coding (or packing). The {\em maximum rate of coding } is denoted by $R$. Indeed people are interested 
to know the asymptotic form of $R$ when $n,d \rightarrow \infty$ and $\delta\equiv d/n$ remains constant. 
So far there is a considerable difference between the best lower and upper bounds for $R$ \cite{H-bell-1950,G-bell-1952,V-akad-1957,MRRW-inform-1977, PL-inform-1998,BJ-codes-2001,S-jcomb-2001,JV-inform-2004}. This means that for large $n$ the best lower and upper bounds for the number of symbols differ by a factor of order $2^n$.

Physically, one can consider the set of symbols as a system of identical particles in the Hamming space of dimension $n$, interacting by a hard core potential of range $d$ \cite{PS-jstatphys-1999,PZ-jstatphys-2006}.
The aim is  then to study the physical states of different densities.  Clearly for small densities the system is in the liquid phase respecting the translational symmetry. In this case it is easy to calculate, for example, the entropy.  At higher densities the liquid entropy becomes incorrect (negative) signaling the onset of other stable phases, either crystalline or glassy  \cite{PZ-jchemphys-2005,PZ-jstatmech-2006}.

In this study we formulate the packing problem as a {\em constraint satisfaction problem}.  We consider $N$ variables (the physical particles or the strings of symbols) which 
take values in $\Lambda$ and for each pair of the variables we consider a constraint that forbids overlapping assignments of the two variables.
A packing is thus an assignment of the variables that satisfies all the constraints.  This representation differs substantially with the so called  {\em lattice gas} models where binary variables (representing occupied or empty positions) are defined one each point of the space $\Lambda$.  

The {\em cavity method } provides analytical and numerical tools which can be extremely useful in solving optimization problems (or constraint satisfaction problems)  over random structures \cite{MP-epjb-2001,MPZ-science-2002,MZ-pre-2002,MP-jstatphys-2003,MPV-1987}. 
In certain cases  it is known to provide sampling results which cannot be obtained by Monte Carlo Markov chains in subexponential times [e.g. optimization problems in the one-step replica symmetry breaking (1RSB) phase]. As an optimization tool it often outperforms linear programming methods \cite{PNAS-nosotros-2011}. 
The development of such algorithm is, however, by no means obvious due to  the choice of the representation of the problem and to the need of writing the cavity equations in an algorithmically efficient form.   

As we shall discuss in this paper, insights from the application of the cavity method  will turn out to be also useful in the study of the type of packing problems we are interested in.

In this paper we study the cavity equations for the packing problem in the replica symmetric (RS) and in the one-step replica symmetry breaking (1RSB) approximations \cite{MPV-1987}. 
These equations are called {\em belief propagation } (BP) and {\em survey propagation } (SP) equations,
respectively \cite{MZ-pre-2002,BMZ-rsa-2005}.  In the RS approximation, besides a liquid solution we find a crystalline phase
where with higher probability spheres are found in one of the sublattices (even or odd) of the Hamming space. This phase has already been observed in Monte Carlo simulations of Ref. \cite{PZ-jstatmech-2006}. Both the liquid and crystalline solutions predict a maximum rate of packing that behaves asymptotically like the best known lower bound \cite{JV-inform-2004}. 
The same result has been obtained in Ref. \cite{PZ-jstatphys-2006} where the  liquid entropy is computed in  the hypernetted chain approximation.
To discuss the exactness of the  cavity free entropy we also consider some interpolation techniques which connect the cavity free entropy to the true entropy of the system \cite{G-commath-2003,FL-jstatphys-2003,FLT-jstatphys-2003}.

In the 1RSB case, we provide an approximate solution of the SP equations to calculate the configurational entropy,  defined as the (log of)  the number of pure states or clusters in the solution space. This quantity acquires a nonzero value at a clustering transition where the liquid entropy is still positive. 
The maximum rate of packing that is achievable still coincides with the one obtained in the RS approximation.

Finally we design a  message passing algorithm based on the BP equations which allows us to find dense packings in not too large  dimensions.
These packings are typically hard to find  by other simple methods like  Monte Carlo based algorithms. 
Unfortunately the computation time and memory increase  exponentially with the space dimension $n$, making larger dimensions very difficult to explore. To partially overcome this problem we introduce an approximate update rule for the message passing algorithm which is restricted to a subspace and makes the computation more efficient.  This improvement, together with the distributive nature of the algorithm, could help to run the algorithm in larger dimensions.  We also discuss another iterative algorithm that, given a packing configuration of spheres with diameter $d$, finds another packing
for larger diameter $d+1$ by increasing the space dimension $n$. This is a polynomial time algorithm generating maximum packings in an ultrametric space. When applied to our problem, we find crystalline packings predicted by the RS cavity equations for even sphere diameters.

The structure of the paper is as follows. In Sec. \ref{1} we define the problem more precisely and give a summary of known results. 
In Secs. \ref{2} and \ref{3} we present the BP and SP equations and study their consequences for the hard sphere packing problem.
In Sec. \ref{4} we study the packing algorithms and Sec. \ref{5} is devoted to extension to the $q$-ary Hamming spaces. Finally the concluding remarks are given in Sec. \ref{6}. 
In the first two appendixes we give the details of calculations for checking the stability of the BP solutions 
and deriving the SP equations. The interpolation methods and some of their properties are presented in Appendix \ref{inter-app}.

\section{Definitions and known results}\label{1}
Consider $N$ hard spheres of diameter $d$ indexed by $i=1,\ldots,N$ and a Hamming space of dimension $n$. The set of points in this space are denoted by
$\Lambda= \{0,1\}^n$ with size $|\Lambda|=2^n$. We index the points in this space by $\sigma$. A point can be represented by a binary vector of $n$ elements $\in \{0,1\}$. 
The Hamming space can be partitioned into two subspaces: even and odd. Points in the even subspace have an even number of $1$'s (even parity) 
and those in the odd subspace have an odd number of $1$'s (odd parity). The Hamming distance $D(\sigma,\sigma')$ between two points $\sigma$ and $\sigma'$ is equal to the number of different elements in the binary representation of the two points. 
For each point $\sigma$ we define the set $V_d(\sigma)$ as
\begin{equation}
 V_d(\sigma)\equiv \{\sigma'|D(\sigma,\sigma') < d\}, \hskip1cm
V_d\equiv |V_d(\sigma)|=\sum_{l=0}^{d-1} \left(\begin{array}{c}
n\\
l
\end{array}\right).
\end{equation}

The aim is to find a non-overlapping configuration of spheres such that $D(\sigma_i,\sigma_j)\ge d$, for any two spheres $i$ and $j$.
The above problem is a constraint satisfaction problem with
$N(N-1)/2$ constraints to satisfy. We index these constraints by
$(ij)$. A configuration $\underline{\sigma}\equiv \{\sigma_i|i=1,\ldots,N\}$ that satisfies all the constraints is called a solution of the problem. 
The partition function $Z$ counts the number of such solutions,
\begin{equation}
Z= \sum_{\underline{\sigma}} \prod_{i<j} I_{ij}(\sigma_i,\sigma_j),
\end{equation}
where $I_{ij}(\sigma_i,\sigma_j)$ is an indicator function for constraint $(ij)$; it is $1$ if $D(\sigma_i,\sigma_j)\ge d$ and $0$ otherwise.
The maximum rate of packing is defined as
\begin{equation}\label{Rdef}
R \equiv \lim_{n,d\rightarrow \infty} \frac{1}{n} \log_2(N_{max}),
\end{equation}
with $\delta= d/n=\mathrm{const}$ and $N_{max}$ is the maximum number of spheres such that $Z>0$. 

Let us here mention some known lower and upper bounds for $R$.
The Gilbert and Varshamov (GV) lower bound \cite{G-bell-1952,V-akad-1957} states that
\begin{equation}
N_{max}\ge \frac{2^n}{V_d},
\end{equation}
resulting in the following lower bound for the maximum rate of packing
\begin{equation}
R\ge R^{GV}\equiv 1-H(\delta),
\end{equation}
where
\begin{equation}
H(\delta)\equiv -\delta \log_2 \delta -(1-\delta)\log_2 (1-\delta).
\end{equation}
Notice that $1-H(\delta)$ is also the Shannon rate for a binary symmetric channel with error probability $\delta$.
A better lower bound is obtained with graph theoretical methods \cite{JV-inform-2004} and gives
\begin{equation}
R\ge R^{JV}\equiv 1-H(\delta)+\frac{\log_2[c n H(\delta)]}{n},
\end{equation}
where $c$ is a constant. As far as we know, this is the best lower bound reported for the maximum rate of packing. However, it still  
behaves asymptotically like $R^{GV}$. The authors in Ref. \cite{PZ-jstatphys-2006} use the liquid entropy of the system of hard spheres to find a maximum rate of packing that is very close to the above lower bound
\begin{equation}
R^{PZ}\equiv 1-H(\delta)+\frac{\log_2[(2\ln 2) n H(\delta)]}{n}.
\end{equation}

\begin{figure}
\includegraphics[width=10cm]{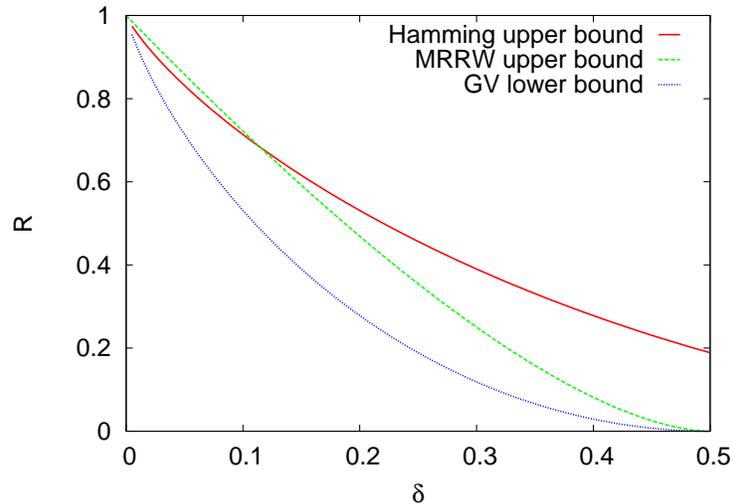}
\caption{Comparing the known lower and upper bounds for the maximum rate of packing $R$.}\label{f1}
\end{figure}

On the other side, we have the Hamming upper bound \cite{H-bell-1950}
\begin{equation}
N_{max}\le \frac{2^n}{V_{\frac{d}{2}}},
\end{equation}
resulting in 
\begin{equation}
R\le R^{H}\equiv 1-H\left(\frac{\delta}{2}\right),
\end{equation}
The best upper bound for $R$ is obtained by the linear programming methods \cite{MRRW-inform-1977}.
An overestimate of this linear programming bound is  
\begin{equation}
R\le R^{MRRW}\equiv H\left(\frac{1}{2}-\sqrt{\delta(1-\delta)}\right).
\end{equation}
In Fig. \ref{f1} we have compared the above bounds to show the large gap between the best lower and upper bounds.

\section{Replica symmetric solutions: BP equations}\label{2}
The basic objects in the cavity method are the cavity marginals or messages passed along the edges of the interaction graph \cite{MP-epjb-2001,MZ-pre-2002,MP-jstatphys-2003}.
In our problem the messages are $2^n$-component vectors with
elements $\in \{0,1\}$. There are two kinds of messages: 
(i) The cavity bias $\mathbf{u}_{i\rightarrow j}$ represents the warning that variable $i$ sends to $j$; $u_{i\rightarrow j}^{\sigma}=1(0)$ 
says that point $\sigma \in \Lambda$ is (not) forbidden by $i$ for variable $j$.
(ii) The cavity field $\mathbf{h}_{i\rightarrow j}$ is sum of warnings that $i$ receives in absence of $j$;
$h_{i\rightarrow j}^{\sigma}$ gives the number unsatisfied constraints, in the absence of $j$, if variable $i$ takes state $\sigma$.
The cavity biases are
\begin{equation}
\mathbf{u}_{i\rightarrow j}\in \{\mathbf{e}_{\sigma}|\sigma \in \Lambda\}, \hskip1cm
e_{\sigma}^{\sigma'}=\left\{
                   \begin{array}{ll}
                     1, & \hbox{if $\sigma' \in V_d(\sigma)$;} \\
                     0, & \hbox{otherwise.}
                   \end{array}
                 \right.
\end{equation}

In the RS framework we assume that all the packings or solutions belong to the same cluster of solutions in the configuration space.
Suppose the interaction graph is a tree and $\underline{\sigma}^*$ is a solution of the problem in the absence of variable $j$. 
Then we represent the cavity bias $\mathbf{u}_{i\rightarrow j}$ by $\mathbf{e}_{\sigma_i^*}$.
Notice that according to our definitions a cavity solution uniquely determines the cavity message $\mathbf{u}_{i\rightarrow j}$.
The histogram of cavity biases among the solutions is given by
\begin{eqnarray}
Q_{i\rightarrow j}(\mathbf{u})=\sum_{\sigma}\eta_{i\rightarrow
j}^{\sigma}\delta_{\mathbf{u},\mathbf{e}_{\sigma}}.
\end{eqnarray}
Assuming a tree interaction graph one can write equations governing the cavity probabilities $\eta_{i\rightarrow j}^{\sigma_i}$:
\begin{eqnarray}\label{BP0}
\eta_{i\rightarrow j}^{\sigma_i}\propto \prod_{k\in V(i) \setminus j}\left( \sum_{\sigma_k}I_{ik}(\sigma_i,\sigma_k)\eta_{k\rightarrow i}^{\sigma_k} \right),
\end{eqnarray}
where $V(i)$ denotes the set of variables interacting with variable $i$. 
The above equations are called BP equations \cite{KFL-inform-2001,BMZ-rsa-2005}. 

In the Bethe approximation the entropy density of the system is written as
\begin{equation}
s=\frac{1}{N} \left[\sum_i \Delta s_i-\sum_{i<j}\Delta s_{ij}\right],
\end{equation}
where $\Delta s_i$ and $\Delta s_{ij}$ are the entropy shifts by adding
variable $i$ and interaction $(ij)$, respectively. For these quantities we have
\begin{align}\label{ds0}
e^{\Delta s_i} &=\sum_{\sigma}\prod_{j\in V(i)}\left(\sum_{\sigma'} I_{ij}(\sigma,\sigma')\eta_{j\rightarrow i}^{\sigma'} \right)\equiv Z_i,\\ 
e^{\Delta s_{ij}} &=\sum_{\sigma,\sigma'}I_{ij}(\sigma,\sigma')\eta_{i\rightarrow j}^{\sigma}\eta_{j\rightarrow i}^{\sigma'}\equiv Z_{ij}.
\end{align}
In words, $Z_i$ is the probability that variable $i$ can occupy at least one point in the Hamming space and
$Z_{ij}$ is the probability that interaction $(ij)$ is satisfied.

Using the above equations for hard spheres we obtain
\begin{equation}\label{BP}
\eta_{i\rightarrow j}^{\sigma}=\frac{\prod_{k\in V(i) \setminus j}\left(1-\sum_{\sigma' \in V_d(\sigma)}\eta_{k\rightarrow i}^{\sigma'} \right)}
{\sum_{\sigma'}\prod_{k\in V(i) \setminus j}\left(1-\sum_{\sigma'' \in V_d(\sigma')}\eta_{k\rightarrow i}^{\sigma''} \right)},
\end{equation}
and
\begin{align}\label{ds}
Z_i =\sum_{\sigma}\prod_{j\in V(i)}\left(1-\sum_{\sigma' \in V_d(\sigma)}\eta_{j\rightarrow i}^{\sigma'} \right),\hskip1cm
Z_{ij} =1-\sum_{\sigma,\sigma':D(\sigma,\sigma')< d}\eta_{i\rightarrow j}^{\sigma}\eta_{j\rightarrow i}^{\sigma'}.
\end{align}
Now we can try different solutions to the BP equations.

\subsubsection{Liquid solution}\label{21}
\begin{figure}
\includegraphics[width=10cm]{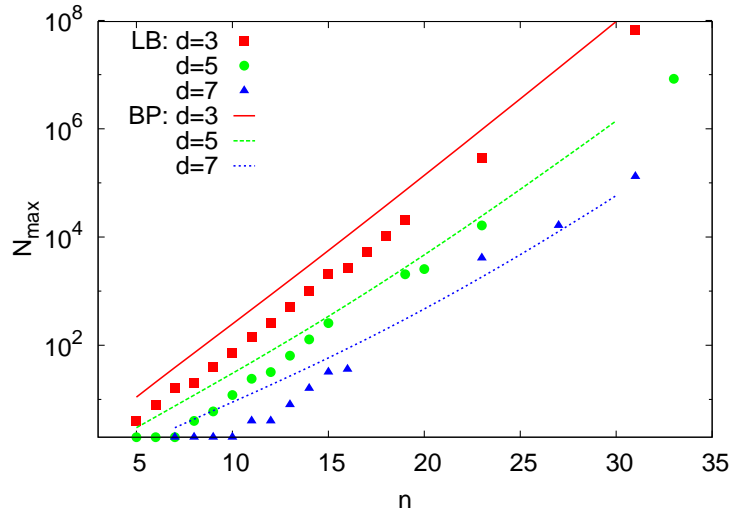}
\caption{Comparing $N_{max}^{BPL}$ (BP) with some lower bounds (LB) from 
\cite{S-amer-1977}. Up to $n=15$ all the lower bounds are exact.}\label{f2}
\end{figure}

The liquid solution is obtained by taking $\eta_{i\rightarrow j}^{\sigma}=\eta=\frac{1}{2^n}$ for any $i$ and $j$.
Evaluating $Z_i$ and $Z_{ij}$ at the liquid solution we can write the BP entropy
\begin{equation}\label{sBP}
s=\ln(2^n)+\frac{N-1}{2}\ln(1-v_d),
\end{equation}
where $v_d\equiv \frac{V_d}{2^n}$.
This entropy vanishes at
\begin{equation}
N_{max}^{BPL}=1-\frac{2\ln(2^n)}{\ln(1-v_d)}.
\end{equation}
For large $n$ it gives
\begin{equation}\label{NmaxBP}
N_{max}^{BPL}v_d \simeq  (2\ln 2) n.
\end{equation}
In Fig. \ref{f2} we compare this quantity with some known exact results and lower bounds in small dimensions.
The maximum rate of packing predicted by the liquid solution of the BP equation is
\begin{equation}
R^{BPL}\simeq 1-H(\delta)+\frac{\log_2[(2\ln 2)n]}{n}.
\end{equation}
where we have used
\begin{equation}
v_d \simeq 2^{-n[1-H(\delta)]}.
\end{equation}
We check the stability of the liquid solution in Appendix \ref{sBP-liquid-app};
we find no continuous glass transition as long as
\begin{equation}
2\ln 2 < \frac{1}{4v_d},
\end{equation}
which is the case for $\delta<\frac{1}{2}$ as $v_d$ is exponentially small in this region. 

\subsubsection{Crystalline solution}\label{22}
In some situations the spheres may prefer one of the sublattices of the Hamming space to the other \cite{PZ-jstatphys-2006}. 
Indeed for even values of $d$, a sphere at the origin forbids more points from the odd sublattice than the even one. 
For high densities the neighboring spheres will be at distance $d$ from the origin, i.e., they occupy points that again belong to the even sublattice. 
The above observation suggests that we could have ordered states in which nearly all the spheres are in the even or odd sector of the Hamming space.  
Here we will consider this situation by assigning different messages $\eta^e$ and $\eta^o$ to points in the even and odd sublattices, respectively. 
Again we use the fact that all the points in one sublattice are equivalent.
Now the BP equations are
\begin{eqnarray}\label{BPcristal}
\eta^e=\frac{(1-V^{e}_d \eta^e-V^{o}_d \eta^o)^{N-2}}{2^{n-1}[(1-V^{e}_d \eta^e-V^{o}_d \eta^o)^{N-2}+(1-V^{e}_d \eta^o-V^{o}_d \eta^e)^{N-2}]},\\ \nonumber
\eta^o=\frac{(1-V^{o}_d \eta^e-V^{e}_d \eta^o)^{N-2}}{2^{n-1}[(1-V^{e}_d \eta^e-V^{o}_d \eta^o)^{N-2}+(1-V^{e}_d \eta^o-V^{o}_d \eta^e)^{N-2}]},
\end{eqnarray}
where $V^{e}_d$ and $V^{o}_d$ are given by
\begin{eqnarray}
V^{e}_d\equiv \sum_{l=0}^{d-1}\frac{1+(-1)^l}{2}\left(\begin{array}{c}
n\\
l
\end{array}\right), \hskip1cm
V^{o}_d\equiv \sum_{l=0}^{d-1}\frac{1-(-1)^l}{2}\left(\begin{array}{c}
n\\
l
\end{array}\right).
\end{eqnarray}
Notice that $\eta^e > 0$ and $\eta^o=0$ cannot be a solution of the above equations unless for $d=n$. 
Indeed to have such a solution we need $\eta^e=\frac{1}{2^{n-1}}=\frac{1}{V^o_d}$, which holds only when $d=n$. 
We will look for other solutions where $\eta^e>0, \eta^o> 0$.
In this case we have
\begin{equation}
\eta^e (1-V^{o}_d \eta^e-V^{e}_d \eta^o)^{N-2} = \eta^o (1-V^{e}_d \eta^e-V^{o}_d \eta^o)^{N-2}.
\end{equation}

\begin{figure}
\includegraphics[width=10cm]{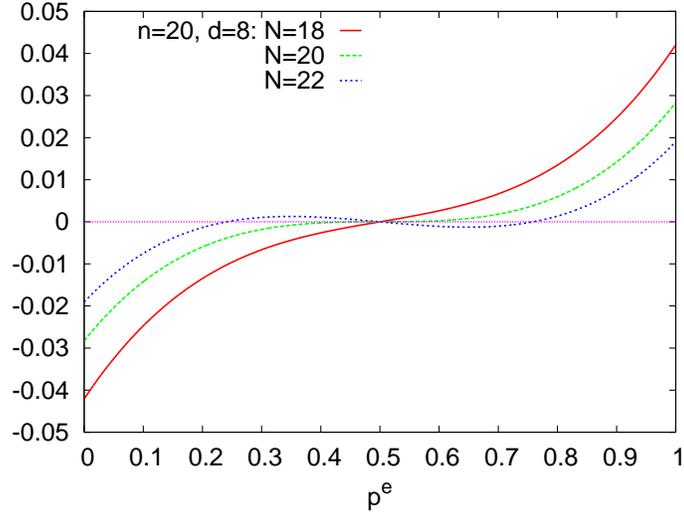}
\caption{The difference between the two sides of Eq. \ref{crys-equ} shows that a crystal solution appears by increasing the number of spheres.}\label{f3}
\end{figure}

\begin{figure}
\includegraphics[width=10cm]{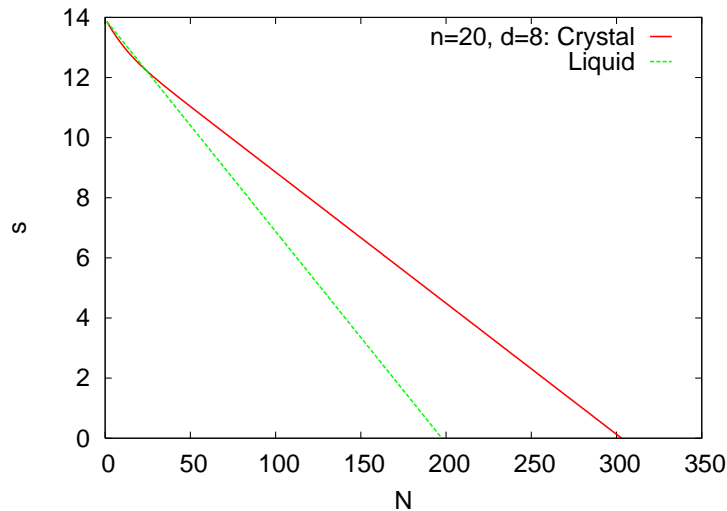}
\caption{Comparing the liquid and crystal entropies predicted by the BP equations.}\label{f4}
\end{figure}

Let us introduce the rescaled variables $v^e_d \equiv V^e_d/2^{n-1}$, $v^o_d \equiv V^o_d/2^{n-1}$, $p^e\equiv 2^{n-1}\eta^e$ and $p^o\equiv 2^{n-1}\eta^o$.
Using the normalization condition $p^e+p^o=1$, we obtain an equation for $p^e$,
\begin{equation}\label{crys-equ}
p^e [1-v^o_d p^e-v^e_d (1-p^e)]^{N-2} = (1-p^e) [1-v^e_d p^e-v^o_d (1-p^e)]^{N-2}.
\end{equation}
For large $n$ and $d$ a nontrivial solution  appears at $N_{0}^{BPC}$; see Fig. \ref{f3}.  
At this point, the two sides of Eq. \ref{crys-equ} have the same slope at $p^e=1/2$.
Thus we find
\begin{equation}
N_{0}^{BPC}=2+\frac{2-\nu_d}{\nu^o_d-\nu^e_d}.
\end{equation}
The entropy of the crystalline phase is obtained given $Z_i$ and $Z_{ij}$,
\begin{align}
Z_i &=2^{n-1}[(1-v^e_d p^e-v^o_d p^o)^{N-1}+(1-v^e_d p^o-v^o_d p^e)^{N-1}],
\\ 
Z_{ij} &=1-p^e(v^e_d p^e+v^o_d p^o)-p^o(v^e_d p^o+v^o_d p^e).
\end{align}
To get a simple expression for the entropy we do the following approximation: $p^e=1$ and $p^o=0$.
Notice that when $d$ is even,  $v^o_d > v^e_d$ and from Eq. \ref{crys-equ} we see that for large $N$, $p^o$ should be much smaller than $p^e$. 
In this approximation the entropy reads 
\begin{eqnarray}
s\approx \ln (2^{n-1})+\frac{N-1}{2}\ln(1-v^e_d).
\end{eqnarray}
This entropy vanishes at
\begin{eqnarray}
N^{BPC}_{max}=1-\frac{2\ln (2^{n-1})}{\ln(1-v^e_d)}.
\end{eqnarray}
For large $n$ the leading term is
\begin{eqnarray}\label{Nmaxcrystall}
N^{BPC}_{max}v^e_d\simeq (2\ln 2) n .
\end{eqnarray}
For $n,d \rightarrow \infty$ the ratio $v_d/v_d^e$ approaches to a constant and we recover asymptotically 
the bound provided with the liquid solution.
In Fig. \ref{f4} we have compared the liquid and crystal entropies for a given $n$ and even $d$. 
Later in Sec. \ref{42} we will introduce an iterative algorithm constructing the above crystalline packings.

\section{One-step RSB solution: SP equations}\label{3}
In the 1RSB framework we assume that there are an exponentially large number $\mathcal{N}_c \sim e^{N\Sigma}$ of clusters of solutions.
A cluster of solutions consists of packing solutions in the configuration space that are connected to each other by paths of finite Hamming distances
in the thermodynamic limit.
In each cluster we could have {\em frozen } and {\em unfrozen } variables.
Suppose the interaction graph is a tree and consider all cavity solutions (in absence of variable $j$) that belong to a given cluster;  
in a tree graph, fixing the boundary variables is equivalent to fixing the cluster of solutions.  
It may happen that in all the solutions of a cluster, variable $i$ takes only one state, say $\sigma$. 
Then the survey bias $\mathcal{U}_{i\rightarrow j}$ will be $\mathbf{e}_{\sigma}$ and 
we say variable $i$ is a completely frozen variable in that cluster. 
We could have partially frozen variables that are frozen on a subset $V^f$ of the Hamming space. 
In this case we represent the survey bias by $\mathcal{U}_{i\rightarrow j}=\sum_{\sigma \in V_{i\rightarrow j}^f} \mathbf{e}_{\sigma}$ 
ignoring the degeneracy of each state. 
In general we could write $\mathcal{U}_{i\rightarrow j}=\sum_{\sigma \in V_{i\rightarrow j}^f} w_{i\rightarrow j}^{\sigma}\mathbf{e}_{\sigma}$ 
where $w_{i\rightarrow j}^{\sigma}>0$ is the number of times that variable $i$ appears in state $\sigma$. 
A variable that takes all possible values is called an unfrozen variable and its survey bias is represented by $\mathbf{0}$.
 
The survey biases defined above depend on the cluster of solutions and change from one cluster to another.
For an edge $(ij)$ of the interaction graph, we define the histogram of the survey biases among the clusters
\begin{eqnarray}
\mathcal{Q}_{i\rightarrow j}(\mathcal{U})= \eta_{i\rightarrow
j}^{0}\delta_{\mathcal{U},\mathbf{0}}+\sum_{m=1}^{2^n-1}\sum_{\sigma_1<\dots<\sigma_m}\eta_{i\rightarrow
j}^{\sigma_1,\dots,\sigma_m}\delta_{\mathcal{U},\sum_{\sigma \in \{\sigma_1,\dots,\sigma_m\}}\mathbf{e}_{\sigma}}.
\end{eqnarray}
Notice that compared with the RS case here we have the extra option $\mathbf{0}$.
The survey fields are determined by the survey biases:
\begin{equation}\label{Hij}
\mathcal{H}_{i\rightarrow j}^{\sigma}=\sum_{k\in V(i) \setminus j} \min_{\sigma'\in V_{k\rightarrow i}^f} \{1-I_{ik}(\sigma,\sigma')\}.
\end{equation}
In words, $\mathcal{H}_{i\rightarrow j}^{\sigma}$ is the minimum number of unsatisfied constraints, in the absence of $j$, if sphere $i$ takes position $\sigma$. Here we need only to know if $\mathcal{H}_{i \to j}^{\sigma}$ is zero or greater than zero.
So we change the field's definition to 
\begin{equation}
\mathcal{H}_{i\rightarrow j}^{\sigma}=\min \left\{1, \sum_{k\in V(i) \setminus j} \min_{\sigma'\in V_{k\rightarrow i}^f} \{1-I_{ik}(\sigma,\sigma')\} \right\}.
\end{equation}
The aim is to write an equation for $\eta_{i\rightarrow j}^{\sigma_1,\ldots,\sigma_m}$. The value of $\eta_{i\rightarrow j}^0$ is determined by the
normalization condition. Variable $i$ sends the survey bias  $\mathcal{U}_{i\rightarrow j}=\mathbf{e}_{\sigma_1}+\cdots+\mathbf{e}_{\sigma_m}$ to variable $j$ if
its survey field is
\begin{equation}
\mathcal{H}_{i\rightarrow j}^{\sigma}=\left\{
                             \begin{array}{ll}
                               0, & \hbox{if $\sigma\in \{\sigma_1,\ldots,\sigma_m\}$;} \\
                               1, & \hbox{otherwise,}
                             \end{array}
                           \right.
\end{equation}
that is if the only choices for variable $i$ are the states in $V^f= \{\sigma_1,\ldots,\sigma_m\}$.
We write this probability as
\begin{equation}
1-\mathrm{Prob}\left(\bigcup_{\sigma \in V^f} \mathcal{H}_{i\rightarrow j}^{\sigma}=1 \hskip2mm \mathrm{.OR.} \hskip2mm 
\bigcup_{\sigma \in \Lambda \setminus V^f }\mathcal{H}_{i\rightarrow j}^{\sigma}=0 \right),
\end{equation}
where $\bigcup_{\sigma}\mathcal{H}_{i\rightarrow j}^{\sigma}=0$ is the event that
at least one of the $\mathcal{H}_{i\rightarrow j}^{\sigma}$ is zero.
We have also the condition that variable $i$ dose not receive
contradictory biases. It means that the cavity field
$\mathcal{H}_{i\rightarrow j}$ should not be equal to $\mathbf{1}\equiv
(1,\ldots,1)$. This probability is given by
$\mathrm{Prob}\left(\bigcup_{\sigma}\mathcal{H}_{i\rightarrow j}^{\sigma}=0\right)$; so we
obtain
\begin{equation}\label{SP0}
\eta_{i\rightarrow j}^{\sigma_1,\ldots,\sigma_m}=\frac{1-\mathrm{Prob}\left(\bigcup_{\sigma \in V^f}\mathcal{H}_{i\rightarrow
j}^{\sigma}=1\hskip2mm \mathrm{.OR.} \hskip2mm \bigcup_{\sigma \in \Lambda \setminus V^f
}\mathcal{H}_{i\rightarrow
j}^{\sigma}=0\right)}{\mathrm{Prob}\left(\bigcup_{\sigma}\mathcal{H}_{i\rightarrow
j}^{\sigma}=0\right)}.
\end{equation}
One can use the standard tools in the probability theory to find the above probabilities in terms of $\eta$'s. 
The result is the so called SP equation and is derived with more details in Appendix \ref{SP-app}.
Here we simplify the analysis by assuming that all partially frozen surveys are zero, 
that is, only completely frozen and unfrozen surveys are taken into account. This would be a reasonable approximation when $N$ approaches $N_{max}$. 
We also expect to obtain a larger complexity within this approximation. In this ansatz the SP equation reads
\begin{equation}\label{SP1}
\eta_{i\rightarrow j}^{\sigma}=\frac{\sum_{m=0}^{2^n-1}(-1)^{m}\sum_{\sigma_1<\dots<\sigma_{m} \in \Lambda \setminus \sigma}\prod_{k \in V(i) \setminus j}
\left(1-\sum_{\sigma' \in V_{\cup}(\sigma,\sigma_1,\dots,\sigma_{m})} \eta_{k\rightarrow i}^{\sigma'}\right)}
{\sum_{m=1}^{2^n}(-1)^{m+1}\sum_{\sigma_1<\dots<\sigma_m}\prod_{k \in V(i) \setminus j}
\left(1-\sum_{\sigma' \in V_{\cup}(\sigma_1,\dots,\sigma_m)} \eta_{k\rightarrow i}^{\sigma'}\right)},
\end{equation}
where $V_{\cup}(\sigma_1,\ldots,\sigma_m)$ is the union volume of $V_d(\sigma_1)$, $V_d(\sigma_2)$, $\ldots$ and  $V_d(\sigma_m)$.

In our problem all the edges of the interaction graph are equivalent. Moreover, due to the translational symmetry, the surveys $\eta_{i\rightarrow j}^{\sigma}$ 
do not depend on $\sigma$. Considering these simplifications, the SP equation can be rewritten in a more compact form as
\begin{equation}\label{SP2}
\eta=\frac{1}{2^n}\frac{\sum_{V=V_d}^{2^n}G'(V)(1-V\eta)^{N-2}}{\sum_{V=V_d}^{2^n}G(V)(1-V\eta)^{N-2}},
\end{equation}
where
\begin{align}\label{Gv}
G(V) = \sum_{m=1}^{2^n}(-1)^{m+1}g_m(V),\hskip1cm
G'(V) =\sum_{m=1}^{2^n}(-1)^{m+1}mg_m(V).
\end{align}
Here $g_m(V)$ is the total number of configurations that $m$ distinct
points can take in the Hamming space such that the union volume $V_{\cup}$ is equal to $V$. More precisely we have
\begin{equation}
g_m(V)=\sum_{\sigma_1<\dots<\sigma_m} \delta_{V,V_{\cup}(\sigma_1,\ldots,\sigma_m)}.
\end{equation}
To obtain more explicit results we consider a naive approximation where  $V_{\cup}(\sigma_1,\ldots,\sigma_m)\approx mV_d$.  
This is a good approximation for small $d/n$. We also approximate $(1-mV_d\eta)^{N-2}$ by $e^{-(N-2)mV_d\eta}$
to compensate the error made by overestimating the union volume $V_{\cup}(\sigma_1,\ldots,\sigma_m)$ for large $m$. Then the SP equation reads
\begin{equation}\label{SP3}
\eta=\frac{1}{2^n}\frac{\sum_{m=1}^{2^n} m (-1)^{m-1}\left(\begin{array}{c}
                                     2^n \\
                                     m
                                   \end{array}\right)
e^{-m(N-2)V_d\eta} }{\sum_{m=1}^{2^n}(-1)^{m-1}\left(\begin{array}{c}
                                     2^n \\
                                     m
                                   \end{array}\right)
e^{-m(N-2)V_d\eta}}.
\end{equation}
Summing over $m$ we get
\begin{equation}\label{SP4}
\eta=\frac{ e^{-(N-2)V_d\eta}(1-e^{-(N-2)V_d\eta})^{2^n-1}
}{1-(1-e^{-(N-2)V_d\eta})^{2^n}}.
\end{equation}
Let us see where the above equation admits a nontrivial solution $\eta > 0$.
To this end we will consider the following scalings: $(N-2)V_d/2^n=c n$ and
$2^n\eta=1-g(n,c)$. For $n\rightarrow \infty$  we expect to have
$c\rightarrow \mathrm{const}$ and $g\rightarrow 0$. Replacing these in the above equation and
expanding for small $2^ne^{-c n(1-g)}$ we obtain
\begin{equation}\label{SPg}
g\simeq 2^ne^{-cn(1-g)}.
\end{equation}
To have a solution for $g$ we need
$c>\ln 2 + c' (\ln n)/n$ with
\begin{equation}
(n\ln 2+c'\ln n)e^{-c' \ln n}=1.
\end{equation}
Increasing $N$, the frozen variables appear for the first time at $N_c^{SP}$, where 
\begin{align}\label{clusterc}
c'\simeq  1+O\left(\frac{1}{\ln n}\right), \hskip1cm
c \simeq \ln 2 +\frac{\ln n}{n}+O\left(\frac{1}{n}\right),
\end{align}
and therefore
\begin{equation}
N_c^{SP}v_d \simeq n\ln 2 +\ln n+\mathrm{const}.
\end{equation}
This gives the clustering transition, when the solution space splits into an exponentially large number of clusters.

\subsubsection{Complexity}\label{31}
We can compute the complexity in the Bethe approximation as 
\begin{equation}\label{complexity}
\Sigma=\frac{1}{N}\left[\sum_i \Delta \Sigma_i-\sum_{i<j}\Delta \Sigma_{ij}\right].
\end{equation}
As before, $\Delta \Sigma_i$ and $\Delta \Sigma_{ij}$ are the complexity shifts by adding
variable $i$ and interaction $(ij)$. 
The complexity is zero for small densities of the particles where we expect to have a single cluster of solutions. 
As we increase $N$, we encounter the clustering transition at $N_c$, where $\Sigma$ becomes nonzero. 
The complexity is a decreasing function of $N$ and finally 
vanishes at $N_{max}$. Here we shall focus on the most numerous clusters, which are not necessarily the relevant ones. 
More accurate results can be  obtained by a large deviation study of the problem, which involves computing 
the complexity of different clusters \cite{KMRSZ-pnas-2007}.

\begin{figure}
\includegraphics[width=10cm]{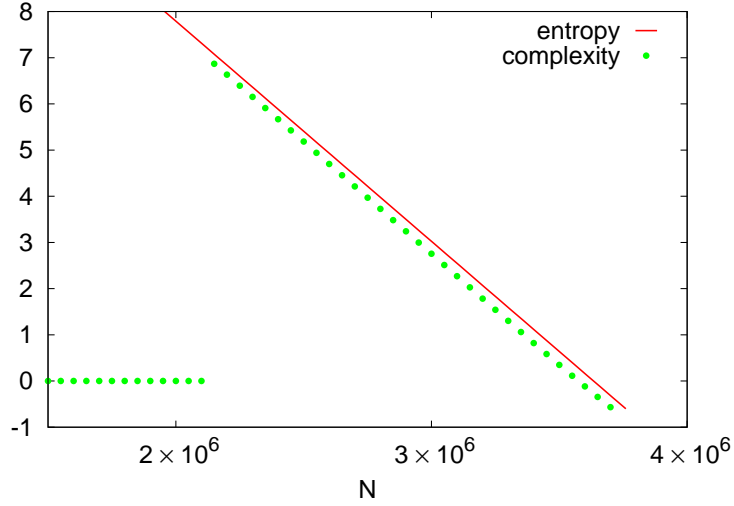}
\caption{Comparing the RS entropy with the 1RSB complexity in the naive approximation. Here $n=25$ and $d=2$.}\label{f5}
\end{figure}

To compute the complexity we need the two quantities $\Delta \Sigma_i$ and $\Delta \Sigma_{ij}$. 
Again for the sake of simplicity we only work with the completely frozen surveys. In this approximation we get
\begin{equation}
e^{\Delta \Sigma_i}=\sum_{m=1}^{2^n}(-1)^{m+1}\sum_{\sigma_1<\dots<\sigma_m}\prod_{j\in
V(i)}\left(1-\sum_{\sigma' \in V_{\cup}(\sigma_1,\ldots,\sigma_m)}\eta_{j\rightarrow i}^{\sigma'}\right)\equiv \mathcal{Z}_i,
\end{equation}
and the link contribution is
\begin{equation}
e^{\Delta \Sigma_{ij}}=1-\sum_{\sigma,\sigma':D(\sigma,\sigma')< d}\eta_{i\rightarrow j}^{\sigma}\eta_{j\rightarrow
i}^{\sigma'}\equiv \mathcal{Z}_{ij}.
\end{equation}
Using our notation in the previous subsection we can write a more
compact form of the complexity for uniform surveys
\begin{eqnarray}\label{complexityc}
\Sigma=\ln \left[\sum_V G(V)(1-V\eta)^{N-1}\right]
-\frac{N-1}{2}\ln[1-2^{n}V_d \eta^2].
\end{eqnarray}
In the naive approximation, where we approximate $V_{\cup}$ by $mV_d$, we find
\begin{equation}\label{Sigman}
\Sigma=\ln[1-(1-e^{-(N-1)V_d\eta})^{2^n}]-\frac{N-1}{2}\ln[1-2^{n}V_d\eta^2].
\end{equation}
The complexity vanishes when
\begin{equation}
1-(1-e^{-(N-1)V_d\eta})^{2^n}=(1-2^{n}V_d\eta^2)^{\frac{N-1}{2}}.
\end{equation}
Using again the scalings $(N-1)V_d/2^n=c n$ and $2^n\eta=1-g(c,n)$, the above equation can be rewritten as
\begin{equation}
1-e^{-x}=e^{-\frac{1}{2}c n(1-g)^2}, \hskip1cm  x\equiv e^{n[\ln 2-c(1-g)]}.
\end{equation}
Expanding the left side for the exponentially small $x$ we obtain
\begin{equation}
2\ln 2=c(1-g^2).
\end{equation}
Notice that $g$ approaches to zero for large $n$ as $1/n$, so the above equation suggests
\begin{equation}\label{NmaxSP}
N_{max}^{SP}v_d= (2\ln 2) n+O(1/n).
\end{equation}
Using Eqs. \ref{SP4} and \ref{Sigman} we can find the complexity in the naive approximation.
In Fig. \ref{f5} we have compared this complexity with the BP entropy for $n=25$ and $d=2$, where the naive
approximation is expected to work.  
We see the jump in the complexity that happens at the clustering transition $N_c^{SP}$. 
Moreover, the complexity is always smaller than the RS entropy, as it should be; it is the sum $\Sigma+s_{cluster}$ that gives the total entropy of the system. 
Here $s_{cluster}$ is the internal entropy of the clusters.

Let us summarize the main approximations we used in the above calculations:
First, the SP equation ignores the soft part of the messages and works only with the hard fields. We know that
this could give an underestimate of the clustering transition but works well in predicting the satisfiability-unsatisfiability (SAT-UNSAT) transition in some constraint satisfaction problems.  
Next, we neglected the partially frozen states and considered only the completely frozen and unfrozen parts to write
a simpler expression for the SP equation. This approximation seems reasonable as for maximum packings the spheres should be strongly localized by their neighborhood.
And finally, we resorted to another approximation by replacing the union volume of $m$ distinct spheres with
$mV_d$. One can improve on this by using an annealed approximation of $V_{\cup}$,
\begin{equation}
V_{\cup}(\sigma_1,\ldots,\sigma_m)\simeq 2^n[1-(1-v_d)^m],
\end{equation}
where $(1-v_d)^m$ is the probability that a point in the Hamming space is outside the $m$ spheres centered at $\sigma_1$, $\ldots$, $\sigma_m$. 
Further, we can approximate $g_m(V)$ by
\begin{eqnarray}
g_m(V)\simeq \left( \begin{array}{c}
 2^n \\
  m
\end{array}\right)
\left( \begin{array}{c}
 2^n \\
  V
\end{array}\right)
[1-(1-v_d)^m]^V[(1-v_d)^m]^{2^n-V}.
\end{eqnarray}
The annealed volume is approximately given by $m V_d$ when $m\ll \frac{2^n}{V_d}$.
Moreover, the completely frozen approximation we used, suggests that the main contribution in the SP equation comes from small values of $m$, see Appendix \ref{SP-app}.
Therefore, we do not expect to observe exponentially large deviations in $N_c^{SP}$ and $N_{max}^{SP}$ beyond the naive approximation.

\section{Some algorithms for the packing problem}\label{4}

\subsection{Hard spheres in an ultrametric space}\label{41}
Exact solutions are always useful in that we obtain some insights about more complex problems \cite{TS-pre-2006,TS-em-2006}. 
As a simple example which can be treated exactly, we consider the packing problem in an ultrametric space.

An ultrametric space is a metric space where for any three points $i_1,i_2,$ and $i_3$ we have $D(i_1,i_2) \le \max \{D(i_1,i_3),D(i_2,i_3)\}$. 
Consider a rooted binary tree $\mathsf{T}_n$ with $n$ generations. The number of points at generation $l$ is $2^l$. 
The space of points is given by the $2^n$ leaves of this tree. The distance $D(i_1,i_2)$ between two points $i_1$ and $i_2$ 
is defined as the number of generations that one should go up in the tree to find the first common ancestor.

\begin{figure}
\includegraphics[width=8cm]{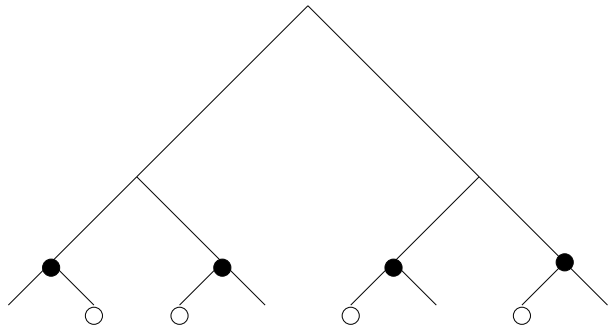}
\caption{Packing hard spheres in an ultra-metric space. Here $n_1=2$ and $N=2^2$. Filled and empty spheres show valid packings for $d=1$ and $d=2$, respectively.}\label{f6}
\end{figure}

The problem is to find a packing of $N$ hard spheres such that for any two spheres $D(i_1,i_2) \ge d$. 
The packing problem is trivial for $d=1$, so we can start from this simple case and try to find packings for
larger $d$. Let us start with a binary tree of $n_1$ generations and $2^{n_1}$ leaves, 
$\mathsf{T}_{n_1}$. We put one sphere at each point of this ultrametric space to obtain a packing of $N=2^{n_1}$ spheres with diameter $d=1$; see Fig. \ref{f6}. 
This is the densest configuration of hard spheres for the above parameters. The idea is to increase $d$ by $1$ and add the minimum number of necessary generations 
to find a new packing with the new diameter for the spheres. Indeed in an ultrametric space we need
only one additional generation to remove all the overlaps in the previous configuration.  After adding the new generation we are free to put each sphere in one of its two descendants. 
One can continue the above process to find a valid packing for an arbitrary value of $d$.
It turns out that these are the densest configurations of the spheres with the parameters $n$ and $d$; the packing indeed partitions the whole space into regions occupied by the spheres. 
Therefore, given $n$ and $d$, the number of packings and maximum number of spheres read
\begin{eqnarray}
Z=N!\left(\begin{array}{c}
2^{n-d}\\
N
\end{array}\right)(2^d)^N, \hskip1cm
N_{max}=2^{n-d}.
\end{eqnarray}
The maximum rate of packing will be
\begin{eqnarray}
R^{UM}=\frac{\log_2 N_{max}}{n}=1-\delta.
\end{eqnarray}
It is interesting that this is exactly the Shannon rate for a binary erasure channel with error probability $\delta$.

The above example is one of the exactly solvable packing problems that one can compare the exact $N_{max}$ with the BP one computed at the 
liquid solution:
\begin{eqnarray}
N_{max}^{BPL}=1-\frac{(2\ln2) n}{\ln(1-2^{d-n})}\to (2 \ln2) n 2^{n-d}.
\end{eqnarray}
Here $N_{max}^{BPL}>N_{max}$ but we find asymptotically $R^{BPL}=1-\delta$.

\subsection{Hard spheres in the Hamming space}\label{42}
Here we can use the same strategy as above to find packings of hard spheres in the Hamming space. 
The points of an $n$-dimensional Hamming space can be represented by the leaves of a binary tree $\mathsf{T}_n$.
A configuration of $N$ spheres, $\underline{\sigma}\equiv \{\sigma_i\in \Lambda|i=1,\ldots,N\}$, is a packing of hard spheres with diameter $d$ 
if it satisfies the following set of constraints
\begin{equation}
\mathcal{C}(d)\equiv \left\{ D(\sigma_i,\sigma_j)\ge d |i\neq j\right\},
\end{equation}
where $D(\sigma_i,\sigma_j)$ is the Hamming distance of points $\sigma_i$ and $\sigma_j$. 
We define the energy $E[\underline{\sigma}]=\sum_{i<j}[1-I_{ij}(\sigma_i,\sigma_j)]$ as the number of unsatisfied constraints. 
The {\em conflict graph } $\mathcal{G}(\underline{\sigma})$ represents a graph of $N$ nodes where edge $(ij)$ is present if the corresponding constraint is not satisfied. 
To find a packing we do the following steps:

\begin{itemize}
\item We start with $N=2^{n_1}$ spheres occupying all the leaves of $\mathsf{T}_{n_1}$.
This is the densest configuration of spheres that satisfies $\mathcal{C}(1)$. We use $\underline{\sigma}^1$ to represent this configuration of spheres.
\item For $t=2,\ldots,d$, we start with $(\underline{\sigma}^{t-1},\mathsf{T}_{n_{t-1}})$ and find $(\underline{\sigma}^{t},\mathsf{T}_{n_{t}})$ such that all the constraints in 
$\mathcal{C}(t)$ are satisfied. Set $n=n_{t-1}$ and $\Delta n_t=0$, then
\begin{itemize}
\item  If $E> 0$, add a new generation to $\mathsf{T}_n$, that is $n=n+1$ and  $\Delta n_t=\Delta n_t+1$.
\item For each sphere find the best value of $\sigma_i(n)\in \{0,1\}$ and call the new configuration $\underline{\sigma}^{t-1,\Delta n_t}$. 
If $E=0$ return $\underline{\sigma}^{t}=\underline{\sigma}^{t-1,\Delta n_t}$ and $n_t=n$.
\end{itemize}
\end{itemize}

\begin{figure}
\includegraphics[width=10cm]{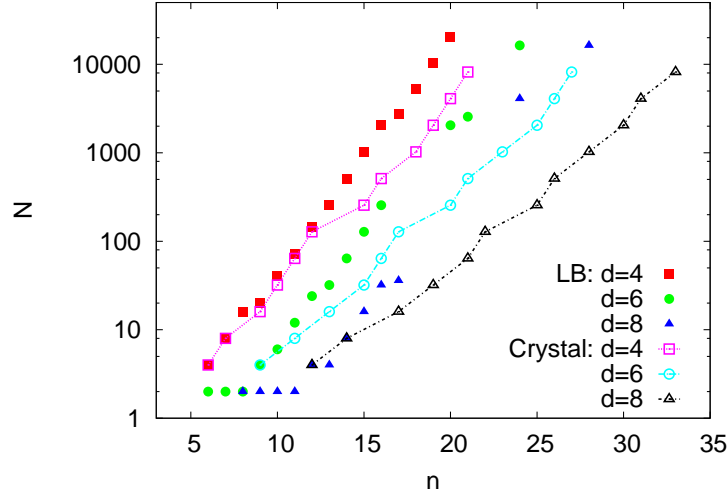}
\caption{Comparing number of spheres in the crystalline packings generated by the iterative algorithm with some lower bounds (LB) from
\cite{S-amer-1977}. Up to $n=15$ all the lower bounds are exact.}\label{f7}
\end{figure}

Notice that every time we add a new generation we have to solve an optimization problem to find the best configuration, 
$\underline{s}\equiv\{\sigma_i(n)|i=1,\ldots,N\}$. Given $t$ and $\Delta n_t$, we need to find the ground state of the following energy function
\begin{equation}\label{Econflict}
E[\underline{s}]= \sum_{(ij) \in \mathcal{G}(\underline{\sigma}^{t-1,\Delta n_t})}\delta_{s_i,s_j}.
\end{equation}
The aim is to assign different values to the new components $s_i$ and $s_j$ when edge $(ij)$ belongs to the conflict graph. In our numerical simulations we use simulated annealing to solve the above optimization problem.

\begin{figure}
\includegraphics[width=10cm]{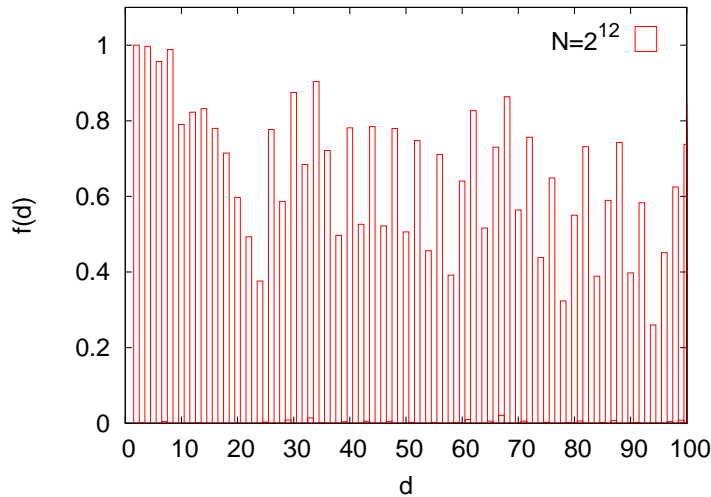}
\caption{Fraction of locally frozen spheres in the packings generated by the iterative algorithm.}\label{f8}
\end{figure}

We observe that for even $t$ the change in the number of generations $\Delta n_t$ is $1$; i.e., the conflict graph is a bipartite graph connecting only spheres in different sublattices, odd and even. 
Consequently, for even $d$ all spheres are in the same sublattice of the Hamming space, as expected from the crystalline phase of Sec. \ref{22}.
In Fig. \ref{f7} we have compared the packings constructed by the above algorithm with the known lower bounds. We also checked the fraction of {\em locally frozen }  spheres in these packings. 
A locally frozen sphere is one that, fixing the other spheres, cannot be locally displaced without violating some of the constraints.
Figure \ref{f8} displays this quantity as a function of $d$. We see that for even $d$ a considerable fraction of the spheres are frozen. This indicates that finding such packings is very difficult by a random sequential adding algorithm \cite{T-springer-2002} where spheres are added randomly one by one.

\subsection{An algorithm based on the BP equations}\label{43}
Let us again write the general BP equations for the packing problem:
\begin{eqnarray}
\eta_{i\rightarrow j}(\sigma_i)\propto \prod_{k\in V(i) \setminus j} \left( \sum_{\sigma_k}I_{ik}(\sigma_i,\sigma_k)\eta_{k\rightarrow i}(\sigma_k) \right).
\end{eqnarray}
To solve these equations one starts with random initial values for the cavity messages and updates the messages iteratively to reach a fixed point.
At the fixed point we have the local marginals given by
\begin{eqnarray}
\eta_{i}(\sigma_i)\propto \prod_{j\in V(i)}\left( \sum_{\sigma_j}I_{ij}(\sigma_i,\sigma_j)\eta_{j\rightarrow i}(\sigma_j) \right).
\end{eqnarray}

One can find a packing configuration by decimation, fixing the spheres one by one according to the above local marginals; after each decimation one has to run again the BP equations
to obtained the new marginals. Here we take another approach, by using the local marginals to converge the equations to a polarized fixed point defining a packing configuration.
We found it useful here to add also some additional random potential $w_i(\sigma_i)$ to break the high level of symmetry present in the problem.
The modified BP equations read 
\begin{align}
\eta_{i\rightarrow j}(\sigma_i) &\propto [\eta_i(\sigma_i)]^{r}e^{\beta w_i(\sigma_i)}\prod_{k\in V(i) \setminus j}
\left( \sum_{\sigma_k}I_{ik}(\sigma_i,\sigma_k)\eta_{k\to i}(\sigma_k) \right),\\ 
\eta_{i}(\sigma_i) &\propto [\eta_i(\sigma_i)]^{r }e^{\beta w_i(\sigma_i)}\prod_{j\in V(i)}
\left( \sum_{\sigma_j}I_{ij}(\sigma_i,\sigma_j)\eta_{j\rightarrow i}(\sigma_j) \right),
\end{align}
where cavity messages are biased toward their local marginals. 
These are the reinforced BP (rBP) equations \cite{BZ-prl-2006} and the positive parameter $r$ is called the reinforcement parameter.
The next step is to write the above rBP equations in the limit $\beta \rightarrow \infty$, assuming the messages scale like $\eta_{i \rightarrow j}(\sigma_i)\propto e^{\beta m_{i\rightarrow j}(\sigma_i)}$.
The resulting reinforced max-sum equations are
\begin{align}
m_{i\rightarrow j}(\sigma_i) &= w_i(\sigma_i)+r m_i(\sigma_i)+\sum_{k\in V(i)\setminus j} \max_{\sigma_k:I_{ik}(\sigma_i,\sigma_k)=1}  m_{k\rightarrow i}(\sigma_k),\\
m_{i}(\sigma_i) &=w_i(\sigma_i)+r m_i(\sigma_i)+\sum_{j\in V(i)} \max_{\sigma_j:I_{ij}(\sigma_i,\sigma_j)=1}  m_{j\rightarrow i}(\sigma_j).
\end{align}

We solve these equations by iteration starting from random initial messages and zero reinforcement.
The reinforcement parameter $r$ is increased gradually as $r(t+1)=r(t)+\delta r$.
At the end, the states that maximize the $m_{i}(\sigma_i)$ define a packing configuration.

The time complexity of this algorithm grows as $[N(2^n-V_d)]^2$. 
We checked the algorithm by finding some packings given in \cite{S-amer-1977}: 
for instance, maximum packings with parameters $(n=10, d=4, N=40)$, $(n=11, d=3, N=144)$, $(n=11, d=5, N=24)$, $(n=15, d=7, N=32)$. 
The main difficulty with large dimensions and number of spheres is the computation time and memory. 
One way to reduce both time and memory is to work with restricted search spaces. For example, 
we may take an initially random search space $\Lambda_i$ of size $S\equiv 2^{n_0}$ for each sphere. 
Given these search spaces we run the reinforced max-sum equations to obtain the cavity and local messages. 
Then for each sphere we replace a part of its search space with new states. The above steps are repeated to find better search spaces and maybe a packing.

More precisely, the algorithm starts by initially randomly selected search spaces $\Lambda_i=\{\sigma_i^{(1)},\dots,\sigma_i^{(S)}\}$. Then we do the following steps: 
\begin{itemize}
\item Run the reinforced max-sum equations: The messages are updated for sufficiently large number of iterations $T$ using the restricted search spaces $\Lambda_i$. 
For example, we use $T=100$ if we increase the reinforcement parameter by $\delta r= 10^{-2}$.
\item Update the search spaces: Compute the local messages $\{m_i(\sigma)|\sigma \in \Lambda_i\}$ and sort them in a decreasing order to put the good states at the beginning of the list. 
Replace a number of states at the end of this list with other states close to the best one. In our simulations we replace half of the search spaces.
\end{itemize}   
In this way we could, for example, find the maximum packings for $(n=10, d=4, N=40)$ and $(n=11, d=3, N=144)$ with a search space of dimension $n_0=3$ and $n_0=5$, respectively. 
This also allowed us to find some packings in larger dimensions, for instance $(n=53, d=29, N=10)$, $(n=54, d=29, N=12)$. These are the best known packings (maybe maximum)
with those parameters. And in some cases we find denser packings with larger sphere diameters: $(n=48, d=32, N=4)$, $(n=51, d=30, N=6)$.

\section{Hard sphere packing in the $q$-ary Hamming space}\label{5}
The above results can easily be extended to the $q$-ary Hamming spaces $\Lambda=\{0,\dots,q-1\}^n$.
There are $q^n$ points in this space represented by vectors of $n$ elements in $\{0,\dots,q-1\}$.
As before, the Hamming distance $D(\sigma,\sigma')$ gives the number of different elements in two vectors 
$\sigma$ and $\sigma'$. Therefore the number of points at distance less than $d$ from a point is given by
\begin{equation}
V_d= |V_d(\sigma)|=\sum_{l=0}^{d-1} \left(\begin{array}{c}
n\\
l
\end{array}\right)(q-1)^l.
\end{equation}
In the asymptotic limit $n,d\to \infty$ and $q,\delta=d/n$ finite, we get $v_d=\frac{V_d}{q^n}\simeq e^{-n[1-H_q(\delta)]}$ where
\begin{eqnarray}
H_q(\delta)=\delta \log_q(q-1)-\delta \log_q(\delta)-(1-\delta) \log_q(1-\delta), \hskip1cm 0\le \delta \le 1-\frac{1}{q}.
\end{eqnarray}

In this section we shall only discuss the results obtained within the replica symmetric approximation.   
The BP equations remain the same as in Eq. \ref{BP}, so we write directly the BP entropy computed at the liquid solution, i.e., $\eta_{i \to j}^{\sigma}=\frac{1}{q^n}$:   
\begin{equation}
s=\frac{N-1}{2}\ln(1-v_d)+\ln(q^n).
\end{equation}
This entropy vanishes at
\begin{equation}
N_{max}^{BPL}=1-\frac{2\ln(q^n)}{\ln(1-v_d)},
\end{equation}
which gives a rate of packing that is not asymptotically different from the Gilbert-Varshamov (GV) one,
\begin{equation}
R^{BPL}\simeq 1-H_q(\delta)+\frac{\log_q[(2\ln q)n]}{n}.
\end{equation}

But, we know that for $q$-ary alphabets and $q$ large enough, the
rate of packing could asymptotically exceed the GV bound
\cite{TV-1991}. These packings, known as algebraic-geometry codes in coding theory, result in the following lower bound when $q$ is a square:
\begin{equation}
R\ge R^{TV}= 1-\delta-\frac{1}{\sqrt{q}-1}, \hskip1cm \hskip1cm 0\le \delta \le 1-\frac{1}{\sqrt{q}-1}.
\end{equation}  
And if $q\ge 49$ there exists always an interval in which $R^{TV}>R^{GV}$.
We believe that these are some crystalline or ordered solutions that only happen for large values of $q$.
The fact that at $\delta=0$ the rate is smaller than $1$ indicates that the above packings are restricted to a subspace
of $\Lambda$ which is exponentially smaller than $q^n$. Moreover, comparing $R^{TV}$ with the rate of packing in an ultra-metric space
suggest that this subspace is effectively ultrametric; see Sec. \ref{41}. More precisely, an ultra-metric 
subspace of size $\simeq q^{n(1-\frac{1}{\sqrt{q}-1})}$ where each sphere occupies $\simeq q^d$ points would result to the same rate of packing as $R^{TV}$. 
We remind that in ultrametric spaces both the BP and GV bounds are asymptotically exact. Notice that also for $q\to \infty$ one obtains $R^{BPL}=R^{GV}=1-\delta$.   

Indeed the number of points that their Hamming distance from a given point is equal to their ultrametric distance is $U_n\equiv \frac{q-1}{q-2}[(q-1)^n-1]$, which is exponentially large as long as $q\ge 3$. This is an upper bound for the size of an ultrametric subspace. Moreover, given diameter $d$, we need only a subspace that is ultra-metric up to distances less than $d$. Let us define  $\mu$ as the probability that a point belongs to an ultrametric subspace. A naive estimation of $\mu$ can be obtained by
\begin{equation}
\mu \simeq (1-\mu)^{V_d-U_d}.
\end{equation}
Note that $U_d$ is exponentially smaller than $V_d$, thus asymptotically $\mu \simeq q^{-nH_q(\delta)}$. In this case, each sphere will occupy only $O(1)$ points of the subspace, so we recover the GV rate of packing. The above argument says the gap between the typical and maximal ultra-metric subspaces is very large, assuming the TV bound comes from packings in an optimal ultrametric subspace.

\section{Conclusion}\label{6}
In summary, we have written the BP and SP equations to study the hard sphere packing problem in the Hamming space. 
Within these approximations we obtained a maximum rate of packing that is asymptotically the same as the lower bound of Gilbert and Varshamov. 
In the RS approximation we also found a crystalline phase where for even values of $d$ the spheres prefer to be in one sector of the
Hamming space. The BP solutions were stable with respect to continuous glass transitions as long as $\delta <1/2$.
This suggests that phase transitions, if any,  would be of discontinues type.  

We have also introduced two new algorithms.
First a  message passing algorithm based on the BP equations  which finds dense packings of hard spheres in finite dimensions.
An approximate scheme is used  to reduce the time and memory complexity of the algorithm, which can still be improved and hopefully lead to new packing results.
An almost identical algorithm can be used in continuous spaces. As a proof of concept we used it to find some known local dense packings in two-dimensional Euclidean space.  

Second,  we introduced an iterative algorithm to find packings of hard spheres starting from small diameters and dimensions.  
For even diameters the algorithm generates packings with all spheres in one subspace of the Hamming space, as expected from
the crystalline solution of the BP equations.

There are still some points that need more effort to be clarified. 
It is of extreme relevance to  establish  the relation between the exact maximum rate of packing and the one provided by the BP equations. 
This means, for instance, that we need to study some interpolating functions between the exact and the BP entropy. Preliminary results are discussed
in Appendix \ref{inter-app}. We expect the BP bound to be asymptotically exact in the binary Hamming space. 
  
Finally,  it would be nice if one could obtain the algebraic-geometry lower bound for the maximum rate of 
packing in the $q$-ary Hamming spaces with some physical arguments.
These are probably crystalline solutions that cannot be captured within the RSB formalism as the RS entropy is expected to be larger than 
that of the glassy solutions in the RSB phase.   

As a concluding remark, we should mention that the 1RSB study presented here is not complete; the SP equations only consider the  
hard or frozen part of the cavity messages. 
A more complete study of the 1RSB approximation also takes the soft part of the messages into account and asks for 
the stability  of these solutions \cite{AR-epjb-2003,RBMM-epjb-2004,APR-jphysa-2004}.

\acknowledgments
We would like to thank C. Baldassi, A. Braunstein, H. Cohn, S. Franz, S. Torquato, and F. Zamponi for their help and useful discussions.
AR and RZ acknowledge the ERC grant OPTINF 267915.

\appendix

\section{Stability of the liquid solution}\label{sBP-liquid-app}
The liquid solution of the BP equations would be stable as long as the ferromagnetic (linear) and spin glass (nonlinear) susceptibilities are finite \cite{AR-epjb-2003,RBMM-epjb-2004,APR-jphysa-2004}. These conditions can be expressed in terms of the maximum eigenvalue of the response matrix $\mathbf{M}$:
\begin{equation}
M_{\sigma,\sigma'}\equiv \frac{\partial \eta_{i\rightarrow j}^{\sigma}}{\partial \eta_{k\rightarrow i}^{\sigma'}}=\left\{ \begin{array}{ll}
    -\frac{1}{2^n}, & \hbox{if $\sigma'\in V_d(\sigma)$;} \\
    \frac{V_d}{2^n(2^n-V_d)}, & \hbox{otherwise.}
  \end{array}
\right.
\end{equation}
that has been evaluated at the liquid solution. Then the stability conditions read
\begin{eqnarray}\label{BPst0}
N_1^{BPL}|\lambda_{max}|=1, \hskip1cm
N_2^{BPL}|\lambda_{max}|^2 =1,
\end{eqnarray}
where $\lambda_{max}$ is the maximum eigenvalue (in absolute value) of $\mathbf{M}$.

More precisely, for $N< N_1^{BPL}$ and $N<N_2^{BPL}$ we would have no continuous phase transition to an ordered and spin glass phase, respectively. 
Notice that these conditions do not exclude discontinuous phase transitions.

The response matrix $\mathbf{M}$ is symmetric with real eigenvalues and orthogonal eigenvectors. Using the translational symmetry by even vectors we write the eigenvectors as
\begin{equation}
u_{\lambda}^{\sigma}=e^{\hat{i}\pi \mathbf{k}.\mathbf{\sigma}},
\end{equation}
where $\mathbf{\sigma}$ is the binary vector representing a point in the Hamming space and $\mathbf{k}$
is a binary wave vector. The eigenvalues are obtained by plugging the above expression in the eigenvalue equation
\begin{equation}
\lambda u_{\lambda}^{\sigma}=-\frac{1}{2^n}\sum_{\sigma' \in V_d(\sigma) }u_{\lambda}^{\sigma'}+\frac{V_d}{2^n(2^n-V_d)}\sum_{\sigma' \in \Lambda \setminus V_d(\sigma) }u_{\lambda}^{\sigma'}.
\end{equation}
Then for the eigenvalues we obtain
\begin{equation}
\lambda=-\frac{1}{2^n}\sum_{\sigma' \in V_d(\sigma) }e^{\hat{i} \pi \mathbf{k}.(\mathbf{\sigma'}-\mathbf{\sigma})}+\frac{V_d}{2^n(2^n-V_d)}\sum_{\sigma' \in \Lambda \setminus V_d(\sigma) }e^{\hat{i} \pi \mathbf{k}.(\mathbf{\sigma'}-\mathbf{\sigma})}.
\end{equation}
The above expression is independent of $\sigma$ and for simplicity we choose $\mathbf{\sigma}=(0,0,\ldots,0)$.
Consider the maximum wave vector $\mathbf{k}=(1,1,\ldots,1)$. The corresponding eigenvalue is
\begin{eqnarray}
\lambda(\mathbf{1})=-\frac{1}{2^n}\sum_{l=0}^{d-1}(-1)^l\left(\begin{array}{c}
n\\
l
\end{array}\right)+\frac{V_d}{2^n(2^n-V_d)}\sum_{l=d}^{n}(-1)^l\left(\begin{array}{c}
n\\
l
\end{array}\right).
\end{eqnarray}
Obviously, the maximum eigenvalue is bounded by $2 v_d$.  
Let us approximate the maximum eigenvalue by the largest contribution at $l=n/2$,
\begin{equation}
|\lambda_{max}|\simeq \frac{V_d}{2^n(2^n-V_d)}\left(\begin{array}{c}
n\\
\frac{n}{2}
\end{array}\right).
\end{equation}
Notice that for large $n$ this is still asymptotically equivalent to the trivial bound $2v_d$. 
Using the Sterling approximation for large $n$ we get
\begin{equation}
|\lambda_{max}|\simeq \frac{2}{\sqrt{n}}\frac{V_d}{(2^n-V_d)},
\end{equation}
which according to Eq. \ref{BPst0} gives
\begin{eqnarray}
N_1^{BPL}v_d\simeq \frac{\sqrt{n}}{2}(1-v_d), \hskip1cm
N_2^{BPL}v_d\simeq \frac{n}{4v_d}(1-v_d)^2.
\end{eqnarray}

\section{Survey propagation equation}\label{SP-app}
Let us start from Eq. \ref{SP0} and find a more suitable form of the SP equation,
\begin{equation}\label{app-SP0}
\eta_{i\rightarrow j}^{\sigma_1,\ldots,\sigma_m}=\frac{1-\mathrm{Prob}\left(\bigcup_{\sigma \in V^f}\mathcal{H}_{i\rightarrow
j}^{\sigma}=1\hskip2mm \mathrm{.OR.} \hskip2mm \bigcup_{\sigma \in \Lambda \setminus V^f
}\mathcal{H}_{i\rightarrow
j}^{\sigma}=0\right)}{\mathrm{Prob}\left(\bigcup_{\sigma}\mathcal{H}_{i\rightarrow
j}^{\sigma}=0\right)},
\end{equation}
where $V_f=\{\sigma_1,\ldots,\sigma_m\}$.
Using the inclusion-exclusion theorem we write
\begin{multline}\label{app-prob0}
\mathrm{Prob}\left(\bigcup_{\sigma}\mathcal{H}_{i\rightarrow
j}^{\sigma}=0\right)=\sum_{\sigma_1}\mathrm{Prob}\left(\mathcal{H}_{i\rightarrow
j}^{\sigma_1}=0 \right)
-\sum_{\sigma_1<\sigma_2}\mathrm{Prob}\left(\bigcap_{\sigma=\sigma_1,\sigma_2}\mathcal{H}_{i\rightarrow
j}^{\sigma}=0 \right)
\\  +\sum_{\sigma_1<\sigma_2<\sigma_3}\mathrm{Prob}\left(\bigcap_{\sigma=\sigma_1,\sigma_2,\sigma_3}\mathcal{H}_{i\rightarrow
j}^{\sigma}=0\right)-\cdots
+(-1)^{2^n-1}\mathrm{Prob}\left(\bigcap_{\sigma}\mathcal{H}_{i\rightarrow
j}^{\sigma}=0\right).
\end{multline}
Here $\mathrm{Prob}\left(\bigcap_{\sigma=\sigma_1,\ldots,\sigma_m}\mathcal{H}_{i\rightarrow
j}^{\sigma}=0\right)$ is the probability of having $\mathcal{H}_{i\rightarrow
j}^{\sigma_1}=0$, $\mathcal{H}_{i\rightarrow
j}^{\sigma_2}=0$,  $\dots$, $\mathcal{H}_{i\rightarrow
j}^{\sigma_m}=0$.

The probability in the numerator can be rewritten in the same way as
\begin{multline}\label{app-num}
\mathrm{Prob}\left(\bigcup_{\sigma \in V^f}\mathcal{H}_{i\rightarrow j}^{\sigma}=1\hskip2mm \mathrm{.OR.} \hskip2mm \bigcup_{\sigma \in \Lambda \setminus V^f}\mathcal{H}_{i\rightarrow j}^{\sigma}=0\right) 
= 
\Bigg\{ \mathrm{Prob}\left(\bigcup_{\sigma \in V^f}\mathcal{H}_{i\rightarrow
j}^{\sigma}=1\right) +\sum_{\sigma_1\in \Lambda \setminus V^f}\mathrm{Prob}\left(\mathcal{H}_{i\rightarrow
j}^{\sigma_1}=0\right) \Bigg\} 
\\ -\Bigg\{ \sum_{\sigma_1\in \Lambda \setminus V^f}\mathrm{Prob}\left(\bigcup_{\sigma \in V^f}\mathcal{H}_{i\rightarrow
j}^{\sigma}=1 \hskip2mm \mathrm{.AND.} \hskip2mm \mathcal{H}_{i\rightarrow j}^{\sigma_1}=0 \right) 
+\sum_{\sigma_1<\sigma_2 \in \Lambda \setminus V^f}\mathrm{Prob}\left(\bigcap_{\sigma=\sigma_1,\sigma_2}\mathcal{H}_{i\rightarrow j}^{\sigma}=0 \right) \Bigg\}+\cdots.
\end{multline}
In the above equation we have probabilities like
\begin{equation}
\mathrm{Prob}\left(\bigcup_{\sigma \in V^f}\mathcal{H}_{i\rightarrow j}^{\sigma}=1\hskip2mm \mathrm{.AND.} \hskip2mm
\bigcap_{\sigma \in \Lambda \setminus V^f}\mathcal{H}_{i\rightarrow j}^{\sigma}=0\right),
\end{equation}
which, using the normalization condition, can be rewritten as
\begin{align}
\mathrm{Prob}\left(\bigcap_{\sigma \in \Lambda \setminus V^f }\mathcal{H}_{i\rightarrow
j}^{\sigma}=0\right)-\mathrm{Prob}\left(\bigcap_{\alpha \in V^f }\mathcal{H}_{i\rightarrow j}^{\sigma}=0\hskip2mm \mathrm{.AND.}
\hskip2mm \bigcap_{\sigma \in \Lambda \setminus V^f }\mathcal{H}_{i\rightarrow j}^{\sigma}=0\right).
\end{align}
Plugging these into our expression for the numerator, Eq. \ref{app-num}, we find
\begin{multline}
\mathrm{Prob}\left(\bigcup_{\sigma \in V^f}\mathcal{H}_{i\rightarrow j}^{\sigma}=1\hskip2mm \mathrm{.OR.} \hskip2mm
\bigcup_{\sigma \in \Lambda \setminus V^f  }\mathcal{H}_{i\rightarrow j}^{\sigma}=0\right)
=
\Bigg\{ \mathrm{Prob}\left(\bigcup_{\sigma \in V^f}\mathcal{H}_{i\rightarrow j}^{\sigma}=1\right)+\sum_{\sigma_1\in \Lambda \setminus V^f}\mathrm{Prob}\left(\mathcal{H}_{i\rightarrow
j}^{\sigma_1}=0\right) \Bigg\} \\ 
-\Bigg\{ \sum_{\sigma_1\in \Lambda \setminus V^f}\mathrm{Prob}\left(\mathcal{H}_{i\rightarrow
j}^{\sigma_1}=0 \right)-\sum_{\sigma_1\in \Lambda \setminus V^f}\mathrm{Prob}\left(\bigcap_{\sigma \in V^f }\mathcal{H}_{i\rightarrow
j}^{\sigma}=0 \hskip2mm \mathrm{.AND.} \hskip2mm \mathcal{H}_{i\rightarrow
j}^{\sigma_1}=0 \right) \\ 
+\sum_{\sigma_1<\sigma_2\in \Lambda \setminus V^f}\mathrm{Prob}\left(\bigcap_{\sigma=\sigma_1,\sigma_2}\mathcal{H}_{i\rightarrow
j}^{\sigma}=0 \right)\Bigg\}+\cdots.
\end{multline}
Notice that the last term in the first bracket is canceled with the first term in the second bracket. This cancellation indeed happens for any two subsequent brackets.  
Simplifying the above expression we find
\begin{multline}\label{app-prob1}
1-\mathrm{Prob}\left(\bigcup_{\sigma \in V^f}\mathcal{H}_{i\rightarrow j}^{\sigma}=1\hskip2mm \mathrm{.OR.} \hskip2mm
\bigcup_{\sigma \in \Lambda \setminus V^f}\mathcal{H}_{i\rightarrow j}^{\sigma}=0\right) \\  
= \mathrm{Prob}\left(\bigcap_{\sigma \in V^f }\mathcal{H}_{i\rightarrow j}^{\sigma}=0 \right)
-\sum_{\sigma_1\in \Lambda \setminus V^f}\mathrm{Prob}\left(\bigcap_{\sigma \in V^f }\mathcal{H}_{i\rightarrow j}^{\sigma}=0
\hskip2mm \mathrm{.AND.} \hskip2mm \mathcal{H}_{i\rightarrow j}^{\sigma_1}=0\right)
\\ 
+\sum_{\sigma_1<\sigma_2 \in \Lambda \setminus V^f}\mathrm{Prob}\left(\bigcap_{\sigma \in V^f }\mathcal{H}_{i\rightarrow
j}^{\sigma}=0 \hskip2mm \mathrm{.AND.} \hskip2mm
\bigcap_{\sigma=\sigma_1,\sigma_2} \mathcal{H}_{i\rightarrow j}^{\sigma}=0
\right)-  
\dots +(-1)^{2^n-|V^f|}\mathrm{Prob}\left(\bigcap_{\sigma} \mathcal{H}_{i\rightarrow
j}^{\sigma}=0\right).
\end{multline}
Now we should write a more explicit relation for
$\mathrm{Prob}\left(\bigcap_{\sigma=\sigma_1,\ldots,\sigma_m}\mathcal{H}_{i\rightarrow
j}^{\sigma}=0\right)$. First note that
\begin{equation}
\mathrm{Prob}\left(\mathcal{H}_{i\rightarrow j}^{\sigma_1}=0\right)=\prod_{k\in
V(i) \setminus j}\left(1-\sum_{m'=1}^{2^n}\sum_{\sigma_1'<\dots<\sigma_{m'}'} I[\sigma_1 \in V_{\cap}(\sigma_1',\dots,\sigma_{m'}')] \eta_{k\rightarrow
i}^{\sigma_1',\dots,\sigma_{m'}'}\right),
\end{equation}
where $V_{\cap}(\sigma_1',\dots,\sigma_{m'}')$ is the intersection of sets $V_d(\sigma_1'),\dots,V_d(\sigma_{m'}')$. 
This is the probability that no variable in $V(i) \setminus j$ sends a
survey that forbids state $\sigma_1$ for variable $i$.
Here $I(\mathcal{C})$ is an indicator function for condition $\mathcal{C}$.
And in general
\begin{equation}\label{app-probcap}
\mathrm{Prob}\left(\bigcap_{\sigma=\sigma_1,\dots,\sigma_m}\mathcal{H}_{i\rightarrow
j}^{\sigma}=0\right) = \prod_{k\in V(i) \setminus j}\left(1-\sum_{m'=1}^{2^n}\sum_{\sigma'_1<\dots<\sigma'_{m'}}^{\cap \{\sigma_1,\dots,\sigma_m\}} \eta_{k\rightarrow
i}^{\sigma'_1,\dots,\sigma'_{m'}} \right),
\end{equation}
where
\begin{equation}
\sum_{\sigma_1'<\dots<\sigma_{m'}'}^{\cap \{\sigma_1,\dots,\sigma_m\}}\equiv \sum_{\sigma_1'<\dots<\sigma_{m'}'} 
I(\{\sigma_1,\dots,\sigma_m\} \cap V_{\cap}(\sigma_1',\dots,\sigma_{m'}') \ne \emptyset ).
\end{equation}
Using the above probabilities in Eqs. \ref{app-prob0} and \ref{app-prob1}, we obtain the SP equation,
\begin{align}
\eta_{i\rightarrow j}^{\sigma_1,\dots,\sigma_m} =  
\frac{\sum_{m'=0}^{2^n-|V^f|}(-1)^{m'}\sum_{\sigma'_1<\dots<\sigma'_{m'} \in \Lambda \setminus V^f}\prod_{k \in V(i) \setminus j}
\left(1-\gamma_{k\to i}(\sigma'_1,\dots,\sigma'_{m'},\sigma_1,\dots,\sigma_m)\right)}
{\sum_{m'=1}^{2^n}(-1)^{m'+1}\sum_{\sigma'_1<\dots<\sigma'_{m'}}\prod_{k \in V(i) \setminus j}
\left(1-\gamma_{k\to i}(\sigma'_1,\dots,\sigma'_{m'}) \right)},
\end{align}
with
\begin{align}
\gamma_{k\to i}(\sigma'_1,\dots,\sigma'_{m'})\equiv \sum_{m''=1}^{2^n}\sum_{\sigma''_1<\dots<\sigma''_{m''}}^{\cap \{\sigma'_1,\dots,\sigma'_{m'}\}} \eta_{k\rightarrow
i}^{\sigma''_1,\dots,\sigma''_{m''}}.
\end{align}

\section{Interpolating between the BP and exact entropies}\label{inter-app}
In any approximation it is important to know how well the method
approximates the correct behavior. Considering the BP approximation, we
know that it is exact at least on tree interaction graphs. On loopy graphs, the BP
approximation may overestimate or underestimate the correct entropy. This in general 
depends on the nature of the interactions and configuration space.

\subsection{A static interpolation}\label{inter-app-1}
The liquid solution of the BP equations is indeed
equivalent to treating all the constraints independently and
replacing a check $I_{ij}(\sigma_i,\sigma_j)$ with its average value
when both $\sigma_i$ and $\sigma_j$ are free. To model this solution,
on each interaction edge $(ij)$ we introduce auxiliary variables
$(s_{ij},s_{ji})$ besides the original variables. The
new variables will serve as independent copies of variables
$(\sigma_i,\sigma_j)$ on edge $(ij)$. Clearly if we want to recover the
exact partition function we should force all the copies to be the
same as their originals. Summing all together we write the following
interpolating partition function:
\begin{eqnarray}
\mathcal{Z}(\beta) \equiv \sum_{\underline{\sigma}}\prod_{i<j}
\left( \frac{1}{(1+e^{-\beta})^{2n}} \sum_{s_{ij},s_{ji}} I_{ij}(s_{ij},s_{ji}) e^{-\beta
[D(\sigma_i,s_{ij})+D(\sigma_j,s_{ji})] }
\right),
\end{eqnarray}
with an inverse temperature $\beta$ scaling as $1/N$.
Then we have $Z^{BPL}=\mathcal{Z}(0)$ and $Z=\mathcal{Z}(
\infty)$. The above interpolating function can be defined for any
constraint satisfaction problem that admits a uniform liquid solution.

Notice that without the normalization factor $1/(1+e^{-\beta})^{2n}$ we would obtain a replicated partition function $\mathcal{Z}_r(\beta)$ 
that is a convex and decreasing function of
$\beta$. In fact $\mathcal{Z}_r(\beta)\ge \mathcal{Z}_r(\infty)=Z$, providing a simple upper pound for the true partition function for any $\beta$. 
To get a good upper bound one needs to solve the
replicated problem for the largest possible $\beta$. Adding the normalization factor destroys the upper bound property but it is necessary if we want to recover $Z^{BPL}$ at $\beta=0$.

Let us define variables $X_i$ as
\begin{equation}
X_i\equiv \sum_{j\neq i} D(\sigma_i,s_{ij}).
\end{equation}
Taking the $m$th derivative of $\ln(\mathcal{Z})$ with respect to
$\beta$ we obtain
\begin{equation}
\frac{\partial^m \ln \mathcal{Z}}{\partial
(-\beta)^m}=\frac{\partial^m \ln
\tilde{Z}}{\partial (-\beta)^m}-
\frac{\partial^m \ln
\tilde{Z}_0}{\partial (-\beta)^m},
\end{equation}
with
\begin{align}
\tilde{Z} =\sum_{\underline{\sigma},\underline{s}}  e^{-\beta
\sum_i X_i} \prod_{i<j} I_{ij}(s_{ij},s_{ji}), \hskip1cm
\tilde{Z}_0 =\sum_{\underline{\sigma},\underline{s}}  e^{-\beta
\sum_i X_i}.
\end{align}
One can easily check that at $\beta=0$ and for $m<6$
\begin{equation}
\frac{\partial^m \ln
\tilde{Z}}{\partial (-\beta)^m}=
\frac{\partial^m \ln
\tilde{Z}_0}{\partial (-\beta)^m}.
\end{equation}
Indeed, it is at $m=6$ that for the first time we encounter
closed loops connecting three spheres in a diagrammatic expansion of the partition function. So for $m<6$ we have
\begin{equation}
\frac{\partial^m \ln \mathcal{Z}}{\partial
(-\beta)^m}|_{\beta=0}=0,
\end{equation}
which means the interpolating partition function is nearly flat
at $\beta=0$. This is also true for $\beta= \infty$ where all
distances $D(\sigma_i,s_{ij})$ should be zero.

\begin{figure}
\includegraphics[width=10cm]{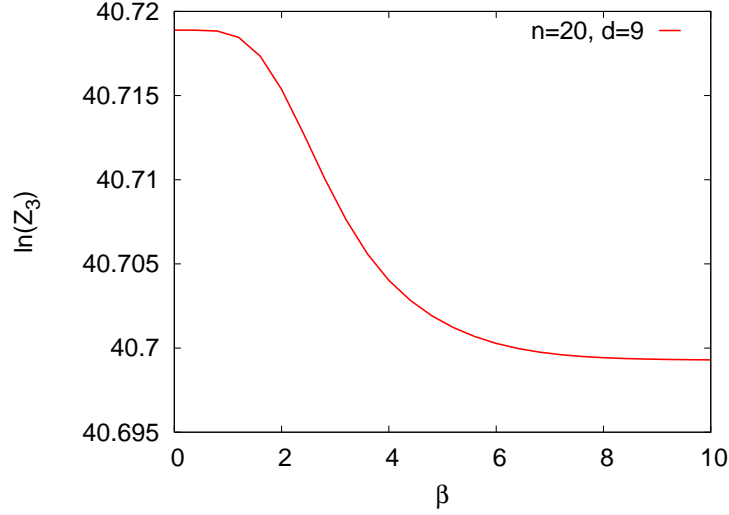}
\caption{The interpolating partition function for three spheres.}\label{f9}
\end{figure}

For $m=6$ we have
\begin{eqnarray}
\frac{\partial^6 \ln \mathcal{Z}}{\partial (-\beta)^6}|_{\beta=0}
= \sum_{i_1<i_2<i_3} \frac{\partial^6 \ln \mathcal{Z}_3}{\partial
(-\beta)^6}|_{\beta=0},
\end{eqnarray}
where $\mathcal{Z}_3$ is the interpolating partition function of
three spheres:
\begin{equation}
\mathcal{Z}_3=\frac{2^n}{(1+e^{-\beta})^{6n}}\sum_{l_{1},l_{2},l_{3}}S_{l_1}Q_{l_2,l_3}(l_1)\prod_{i=1}^3 \left[\sum_{r_3\ge d}\sum_{r_1,r_2} R_{r_1,r_2}(r_3;l_i)e^{-\beta r_1-\beta r_2} \right],
\end{equation}
Here $S_l$ is the number of points at distance $l$ from a given point and 
$Q_{l_1,l_2}(l)$ is the number of points at distance $l_1$ from $\sigma_1$ and distance $l_2$ from $\sigma_2$ when $D(\sigma_1,\sigma_2)=l$.
More precisely, we have
\begin{equation}\label{Ql}
Q_{l_1,l_2}(l)=\left(\begin{array}{c}
n-l\\
\frac{l_1+l_2-l}{2}
\end{array}\right)\left(\begin{array}{c}
l\\
\frac{l+l_1-l_2}{2}
\end{array}\right).
\end{equation}
And finally $R_{r_1,r_2}(r_3;l)$ is the number of pairs $(s_1,s_2)$ that satisfy the following conditions: $D(s_1,\sigma_1)=r_1$, $D(s_2,\sigma_2)=r_2$, and  $D(s_1,s_2)=r_3$ given $D(\sigma_1,\sigma_2)=l$. 
This number can be written as
\begin{equation}
R_{r_1,r_2}(r_3;l)=\sum_r Q_{r_1,r}(l)Q_{r_2,r_3}(r).
\end{equation}
In Fig. \ref{f9} we have plotted
the typical behavior of $\mathcal{Z}_3(\beta)$ for some value of $n$ and $d$. 
Despite the fact that $\mathcal{Z}_3(\beta)$ is decreasing with $\beta$ we find that the first nonzero term in the Taylor expansion
is positive, signaling the nonperturbative nature of $\mathcal{Z}_3(\beta)$ with respect to $\beta$.  

\subsection{A dynamic interpolation}\label{inter-app-2}
Here we use an idea previously introduced in Refs. \cite{FL-jstatphys-2003,FLT-jstatphys-2003} to replace the interactions in the original
problem with the effective ones coming from the BP approximation.
    
Let us label the edges of an interaction graph by $t=1,\dots,M$. Define 
$\mathcal{E}_0$ as the empty set and $\mathcal{E}_t=\{e_1,\dots,e_t\}$ as the set of
the first $t$ edges.
Now we introduce the following sequence of partition functions 
\begin{equation}
Z_t=\sum_{\underline{\sigma}} \prod_{e\in \mathcal{E}_t} I_e(\sigma_{i_e},\sigma_{j_e}).
\end{equation}
Clearly, the original partition function is given by $Z=Z_M$ and \\ $Z^{BPL} = Z_0 \prod_{t=1,M} \langle I_{t}(\sigma_{i_t},\sigma_{j_t}) \rangle_0$ where the average is taken with respect to the uniform and independent distribution of $\sigma_{i_t}$ and $\sigma_{j_t}$.
Suppose we add edge $e_{t+1}$ connecting nodes
$(i_{t+1},j_{t+1})$. Then  
\begin{eqnarray}
Z_{t+1}=Z_{t} \langle I_{t+1} \rangle_0 (1+\Delta_{t+1}),\hskip1cm
\Delta_{t+1}=\frac{ \langle I_{t+1} \rangle_t-\langle I_{t+1} \rangle_0}{\langle I_{t+1} \rangle_0},
\end{eqnarray}
where $\langle I_{t+1} \rangle_t=P_{t}(I_{t+1})$ is the probability of satisfying constraint $I_{t+1}$ when
the interaction set is given by $\mathcal{E}_t$. And $\langle I_{t+1} \rangle_0=P_0(I_{t+1})$ is the same probability when the interaction set is $\mathcal{E}_0$.
Moreover, $\Delta_{t+1}=0$ when the interaction graph
$\mathcal{E}_{t+1}$ is a tree. We rewrite the final entropy as
\begin{eqnarray}\label{corrections}
\log Z=\log Z^{BPL}+\sum_{t=1}^{M} \log(1+\Delta_{t}).
\end{eqnarray}

\begin{figure}
\includegraphics[width=10cm]{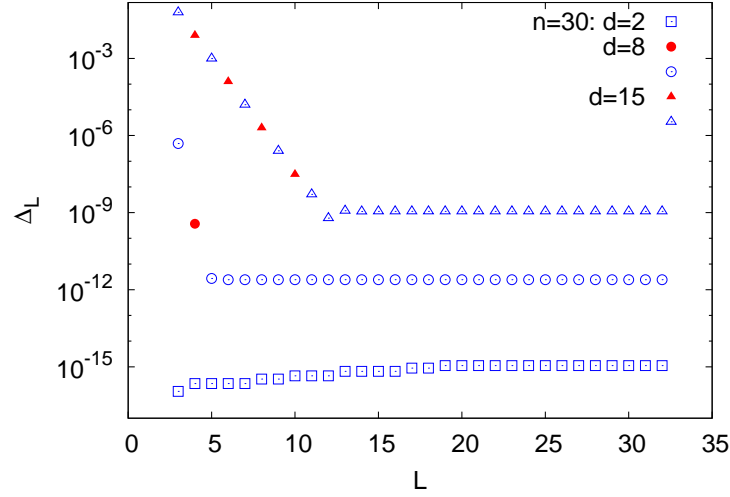}
\caption{Absolute value of $\Delta_L$ in a chain of repulsive interactions. Filled and empty symbols show 
positive and negative values, respectively.}\label{f10}
\end{figure}

For hard spheres $\Delta_t$ may have different signs depending on the interaction graph. For example, suppose
the interaction graph is a chain of size $L$ and we add the interaction between the end points. The probability of finding the end
points at distance $r$ can be obtained in a recursive way as
\begin{eqnarray}
P_{L}(r)=\sum_{r'} P_{L-1}(r') \left( \frac{1}{2^n-V_d} \sum_{r''\ge d} Q_{r,r''}(r')\right),
\end{eqnarray}
using our expression for $Q_{l_1,l_2}(l)$ in Eq. \ref{Ql}.
In this case, having $\Delta_L$ is enough to know if the BP entropy is an over- or underestimation.  
Figure \ref{f10} displays this quantity for different values of $L$.
We observe that $\Delta_L$ is always negative for small $d$, but
for large $d$ and small $L$ its sign alternates as expected for antiferromagnetic interactions.

Looking again at the BP entropy shows that in general the correction in Eq. \ref{corrections}
is relevant only if divided by $N$ it is still exponential in dimension $n$. The corrections are irrelevant for instance if $\Delta_{t}$'s behave as independent random numbers with a symmetric distribution and $\Delta_{t}=O(1)$.
This is what we expect to happen in high dimensions.

It is instructive here to compare the above hard sphere problem with a system of attractive interactions where
the constraints $I_{ij}(\sigma_i,\sigma_j)$ are satisfied if $D(\sigma_i,\sigma_j)\le d$.
Figure \ref{f11} displays $\Delta_L$ in a chain of attractive interactions;
$\Delta_L$ is positive for small $L$ but approaches an exponentially small negative value for large $L$ and $d$.
In this case, we expect to have asymptotically $\Delta_t\ge 0$.  This is similar to the Griffiths inequalities \cite{G-jmath-1967} for ferromagnetic systems. 
Actually, using the Fortuin-Kasteleyn-Ginibre (FKG) inequalities \cite{FKG-cmath-1971} one can prove the above statement for soft attractive interactions, where $I_{ij}(\sigma_i,\sigma_j)=e^{-\frac{\beta}{N} D(\sigma_i,\sigma_j)}$ and
$\beta \ge 0$: 

\begin{figure}
\includegraphics[width=10cm]{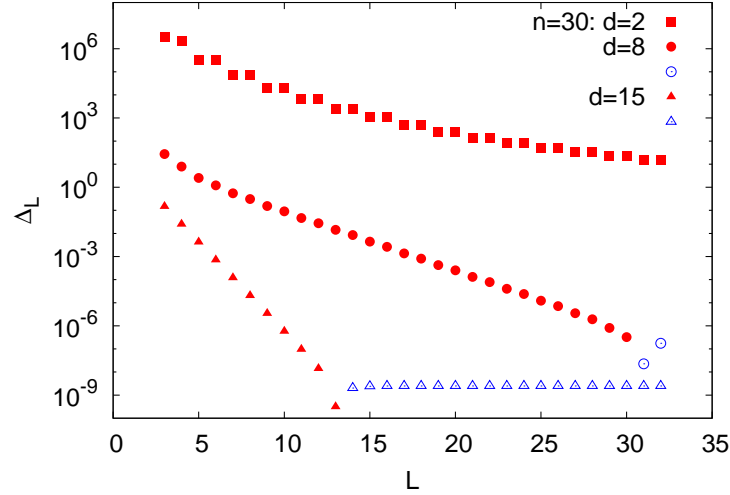}
\caption{Absolute value of $\Delta_L$ in a chain of  attractive interactions. Filled and empty symbols show 
positive and negative values, respectively.}\label{f11}
\end{figure}

Consider the following measure on configurations $\underline{\sigma} \in \Lambda^N$ where $\Lambda=\{0,1\}^n$,
\begin{equation}
\mu(\underline{\sigma})\propto \prod_{e\in \mathcal{E}_t} I_{e}(\sigma_{i_e},\sigma_{j_e}).
\end{equation}
This measure satisfies the conditions of the FKG inequality; $\Lambda^N$ is a partially ordered set
and $\mu(\underline{\sigma})$ is convex, that is given two configurations $\underline{\sigma}$ and $\underline{\sigma}'$   
\begin{equation}
\mu(\underline{\sigma}^{max})\mu(\underline{\sigma}^{min}) \ge \mu(\underline{\sigma})\mu(\underline{\sigma}'),
\end{equation}
where $\underline{\sigma}^{max}=\max(\underline{\sigma},\underline{\sigma}')$ and $\underline{\sigma}^{min}=\min(\underline{\sigma},\underline{\sigma}')$. 
These are bitwise max and min functions. One can use induction to show that
$D(\sigma_i^{max},\sigma_j^{max})+ D(\sigma_i^{min},\sigma_j^{min}) \le  D(\sigma_i,\sigma_j)+ D(\sigma_i',\sigma_j')$ which proves the convexity of $\mu(\underline{\sigma})$
for the soft attractive interactions. 

Having these conditions the FKG inequality says that for decreasing (increasing) functions $f(\underline{\sigma})$ and $g(\underline{\sigma})$ 
\begin{equation}
\langle f g\rangle - \langle f\rangle \langle g\rangle \ge 0,
\end{equation}
where the averages are taken with respect to the measure $\mu(\underline{\sigma})$. A function $f(\underline{\sigma})$ is decreasing if 
$f(\underline{\sigma})\ge f(\underline{\sigma}')$ for any $\underline{\sigma}'\ge \underline{\sigma}$. 
The latter is a bit-wise inequality.

Now consider decreasing functions  $f= I[D(\sigma_i,0)\le l_i] $ and $g= I[D(\sigma_j,0)\le l_j] $ where $0$ is the
zero member of $\Lambda$. According to the FKG inequality we have
\begin{equation}
\langle I[D(\sigma_i,0)\le 0] I[D(\sigma_j,0)\le l] \rangle \ge \langle I[D(\sigma_i,0)\le 0] \rangle \langle I[D(\sigma_j,0)\le l] \rangle,  
\end{equation}
which means $P_t[D(\sigma_i,\sigma_j)\le  l]\ge P_0[D(\sigma_i,\sigma_j)\le l]$ for any $i, j,$ and $l$. In other words, it is more likely to find two spheres closer to each other compared to the uniform measure. As a result, for the soft attractive interactions BP provides a lower bound for the log-partition function.

Actually for the soft interactions $I_{ij}(\sigma_i,\sigma_j)=e^{\pm \frac{\beta}{N} D(\sigma_i,\sigma_j)}$ the partition function reads
\begin{equation}
Z=\sum_{\underline{\sigma}}e^{\pm \frac{\beta}{N}\sum_{i<j} D(\sigma_i,\sigma_j)}=[\sum_{M=0}^N \left(\begin{array}{c}
N\\
M
\end{array}\right) e^{\pm \beta M(1-M/N)}]^n.
\end{equation}
One can exactly solve the problem in the thermodynamic limit and compare the results with the BP predictions to confirm the above statement for the attractive interactions.
In the case of repulsive interactions the BP and exact results asymptotically coincide.

\end{document}